\documentclass[pdflatex,sn-basic]{sn-jnl}

\usepackage{titlesec}
\usepackage{natbib}
\usepackage{enumitem}
\usepackage{tabularx}
\usepackage{booktabs}
\usepackage{hyperref}
\usepackage{bookmark}
\usepackage{graphicx}%
\usepackage{multirow}%
\usepackage{amsmath,amssymb,amsfonts}%
\usepackage{amsthm}%
\usepackage{mathrsfs}%
\usepackage[title]{appendix}%
\usepackage{xcolor}%
\usepackage{textcomp}%
\usepackage{manyfoot}%
\usepackage{booktabs}%
\usepackage{algorithm}%
\usepackage{algorithmic}
\usepackage{listings}%
\usepackage{bookmark}
\usepackage{anyfontsize}

\renewenvironment{table}[1][]%
{\tableorg[#1]%
\tablebodyfont%
\renewcommand\footnotetext[2][]{{\removelastskip\vskip3pt%
\let\tablebodyfont\tablefootnotefont%
\hskip0pt\if!##1!\else{\smash{$^{##1}$}}\fi##2\par}}%
}{\endtableorg}

\usepackage{float}

\newtheorem{theorem}{Theorem}
\newtheorem{lemma}{Lemma}

\begin{document}

\title[Article Title]{Multi-stage Group Testing with (r,s)-regular design Algorithms}

\author*[1]{\fnm{Michael} \sur{Balzer}}\email{michael.balzer@uni-bielefeld.de}

\affil*[1]{\orgdiv{Bielefeld University}, \orgname{Center for Mathematical Economics}, \orgaddress{\street{Universitätsstraße 25}, \city{Bielefeld}, \postcode{33615}, \state{NW}, \country{Germany}}}

\abstract{In industrial engineering and manufacturing, quality control is an essential part of the production process of a product. To ensure proper functionality of a manufactured good, rigorous testing has to be performed to identify defective products before shipment to the customer. However, testing products individually in a sequential manner is often tedious, cumbersome and not widely applicable given that time, resources and personnel are limited. Thus, statistical methods have been employed to investigate random samples of products from batches. For instance, group testing has emerged as an alternative to reliably test manufactured goods by evaluating joint test results. Despite the clear advantages, existing group testing methods often struggle with efficiency and practicality in real-world industry settings, where minimizing the average number of tests and overall testing duration is critical. In this paper, novel multistage $(r,s)$-regular design algorithms in the framework of group testing for the identification of defective products are investigated. Motivated by the application in quality control in manufacturing, unifying expressions for the expected number of tests and expected duration are derived. The results show that the novel group testing algorithms outperform established algorithms for low probabilities of defectiveness and get close to the optimal counting bound while maintaining a low level of complexity. Mathematical proofs are supported by rigorous simulation studies and an evaluation of the performance.}


\maketitle

\section{Introduction} \label{sec:intro}
In various real-world situations, members of a population can be assigned a specific state which often indicates adverse effects on those affected. However, each specific state of a member is usually unknown and subject to uncertainty and ambiguity. Subsequently, the identification of the state becomes a central focus in practice. Members identified with this state can then generally be referred to as defective, while those without it are termed non-defective. For instance, in epidemiology, individuals in a population must be labeled as positive or negative for an infectious disease. Similarly, in industrial engineering, ensuring the proper functionality and quality of manufactured goods is crucial, thereby resulting in labeling of products as either operational or faulty \citep{du1999}.

For the identification of the state, various rigorous testing algorithms have been developed. The most straightforward algorithm involves testing each member separately to determine the presence of the specific state utilizing specifically designed protocols. In principle, the protocol dictates the fundamental approach for conducting individual tests in different areas. For example, in biomedical sciences, such a protocol is the reverse transcription polymerase chain reaction technique \citep{du2006}. Similarly, in manufacturing, quality control protocols ensure that each product is individually inspected for defects \citep{du1999}. However, the individual testing algorithm is often infeasible since time, resources and personnel are generally limited. To overcome these limitations, group testing (GT) has emerged as a practical alternative. In GT, members from a population are grouped and their states are evaluated based on joint test results. The joint test results can then be utilized to infer the condition of the groups. In principle, if a group test in a joint test result returns positive, the corresponding group is labeled as positive which indicates that at least one member in this group is defective. Conversely, members from negative groups can be directly declared as non-defective with a certain level of confidence. However, the specific defective members in a group remain unidentified, thereby requiring the application of additional follow-up strategies. For instance, retesting each member in a positive group separately serves as the foundation of the classical two-stage GT algorithm \citep{dorfman1943}. Alternatively, subsequent stages of GT can be conducted with remaining members of a positive group while the final stage always consists of individual retesting. As testing relies on results from previous stages in these multi-stage GT algorithms, a significant drawback in many applications is the potential delay in identifying defective members \citep{patel1962, li1962}. Due to the reduced number of tests compared to individual testing, GT can lead to significant (economic) cost savings \citep{dorfman1943, hwang1972, du2006}.

Since the advent of the coronavirus disease 2019 pandemic, GT has garnered increased attention with the development of novel GT algorithms which aim to reduce the (economic) cost associated with testing further \citep{aldridge2022b}. Particularly, so-called two-stage $(r,s)$-regular design algorithms (RSA) have been introduced by \cite{aldridge2022b} based on results discussed in \cite{broder2020}. The general algorithmic outline is based on an adjustment of the classical two-stage GT algorithm as introduced in \cite{dorfman1943}. First, in the classical two-stage GT algorithm, all members are evaluated on one joint test result in the first stage. In contrast, in two-stage RSA, all members undergo a number of $r \in \mathbb{N}$ joint tests. To this end, the members in the population are sampled at random $r$ times which are then grouped into groups of size $s \in \mathbb{N}$. Thus, there are $r$ constant number of joint tests per member and $s$ constant number of members per group test, hence the term $(r,s)$-regular design. Second, in contrast to the classical two-stage GT algorithm, only members that are part of a positive group in every joint test $r$ proceed to the second stage of individual retesting. It has been demonstrated that two-stage RSA can substantially reduce the number of tests compared to the classical two-stage GT algorithms, approaching the information-theoretic lower bound. However, the practical implementation of two-stage RSA often proves infeasible in many real-world scenarios. In principle, this infeasibility arises due to the potentially large number of joint tests. Thus, conducting two-stage RSA in practice poses notable challenges since the algorithms require a large, coordinated workforce, extensive effort for additional sample collection and a suitable testing infrastructure that facilitates a large-scale testing procedure \citep{broder2020, mutesa2021, aldridge2022, aldridge2022b}.

While previous research has primarily focused on two-stage RSA and addressing the practical shortcomings by restricting the number of joint tests $r$, a natural extension of two-stage RSA to multiple stages, to the best of my knowledge, has not yet been explored. The aim of this paper is to investigate the behavior of these multi-stage RSA with respect to the expected number of tests (ENT). To this end, expressions for the ENT are derived which is based on the generalization of the expressions provided in \cite{aldridge2022b}. This enables the comparison of established multi-stage GT algorithms and novel multi-stage RSA, which is still missing. Particularly, the performance of multi-stage RSA is evaluated against multi-stage GT algorithms proposed in \cite{patel1962}. The results show that multi-stage RSA generally outperforms established multi-stage GT algorithm for low probabilities of defectiveness. Furthermore,  special cases of the novel multi-stage RSA get very close to the information-theoretical optimal counting bound with a low level of complexity, thereby ensuring practical applicability. Thus, the limitations of two-stage RSA can be effectively mitigated. To illustrate the potential benefits of the novel multi-stage RSA, the earliest practical application of GT in quality control in manufacturing, namely the light bulb testing problem, is revisited. In principle, in the manufacturing industry, ensuring the functionality and quality of products such as light bulbs involves rigorous testing. Light bulbs can usually be arranged in batches. Thus, they can be efficiently grouped, and their functionality evaluated based on joint test results. Unlike many other applications, the rapid availability of test results in these settings makes multi-stage GT algorithms practically feasible \citep{sobel1959, du1999}.

The structure of this paper is as follows: Section \ref{sec:mot} introduces an industrial application involving light bulb testing. In Section \ref{sec:meth}, preliminaries, assumptions, and notations for multi-stage RSA are presented, followed by the derivation of a unifying theorem for the ENT. Afterwards, special cases of multi-stage RSA are discussed. Section \ref{sec:num} provides rigorous simulation studies to support the mathematical proofs, followed by an evaluation of multi-stage RSA. The paper concludes with a summary of the findings and a discussion for future research directions in Section \ref{sec:conc}.

\section{Group Testing for Quality Assurance in Industrial Manufacturing} \label{sec:mot}
The application of statistical methods in quality control procedures in engineering has a long-standing history. For instance, various methods such as ongoing reliability tests, lot quality assurance sampling and acceptance sampling ensure a certain level of quality of manufactured goods. These methods primarily rely on randomly selecting samples from lots or batches, testing the selected items for defects, and then making a decision about the complete corresponding batch or lot from which the items were drawn \citep{lemeshow1991, hobbs2008, nelson2009, gray2016, mittag2018}. In addition to these modern sampling-based approaches, GT has also been employed in quality control scenarios where testing entire batches or pooled samples of products is feasible. One of the earliest described applications of GT in manufacturing involves testing for leaks in devices assembled with chemical substances. After assembly, these devices are placed in a bell jar and tested for the presence of gases. If gas is detected, it indicates that at least one device is leaking and therefore faulty \citep{sobel1959}. More recently, GT has been practically applied in the analysis of the chemical composition of manufactured products. The objective in these cases is to determine whether a substance within a product exceeds a threshold that could potentially induce adverse effects on the user \citep{huang2023}. Similar GT methods are employed in the food and pharmaceutical industries, where batches of food and medicine are tested for harmful chemicals in laboratory settings \citep{herer2000}.

As evidenced by these examples, if manufactured goods can be grouped into batches and evaluated based on joint test results without being destroyed, GT is a feasible and efficient solution. In the following, exemplary schematic schemes of GT algorithms are presented. For this purpose, light bulbs serve as a practical illustrative example. The central goal in this context is to test light bulbs for functionality before they are packaged and shipped. By testing in batches, defective light bulbs can be identified and removed quickly, allowing batches containing only functional light bulbs to be processed efficiently. This approach exemplifies how GT can streamline quality control processes across the electronics industries. \citep{sobel1959, du1999, du2006}. Particularly, four different GT algorithms are presented, each illustrated through an example involving a small population. Specifically, the example consists of a batch of 25 light bulbs, each marked with a unique number. At the beginning of each GT algorithm, the population is randomly sampled. The GT algorithms are then applied, and the final number of tests required is compared across methods. In this example, the defective light bulbs are numbered 1, 3, and 22. The goal is to identify all defective light bulbs with certainty. For simplicity, the group sizes in these examples are chosen arbitrarily. As a reference point, individual testing requires 25 tests to identify all defective light bulbs.

\begin{figure}[H] 
    \centering
    \makebox[\textwidth]{\includegraphics[width=0.8\textwidth]{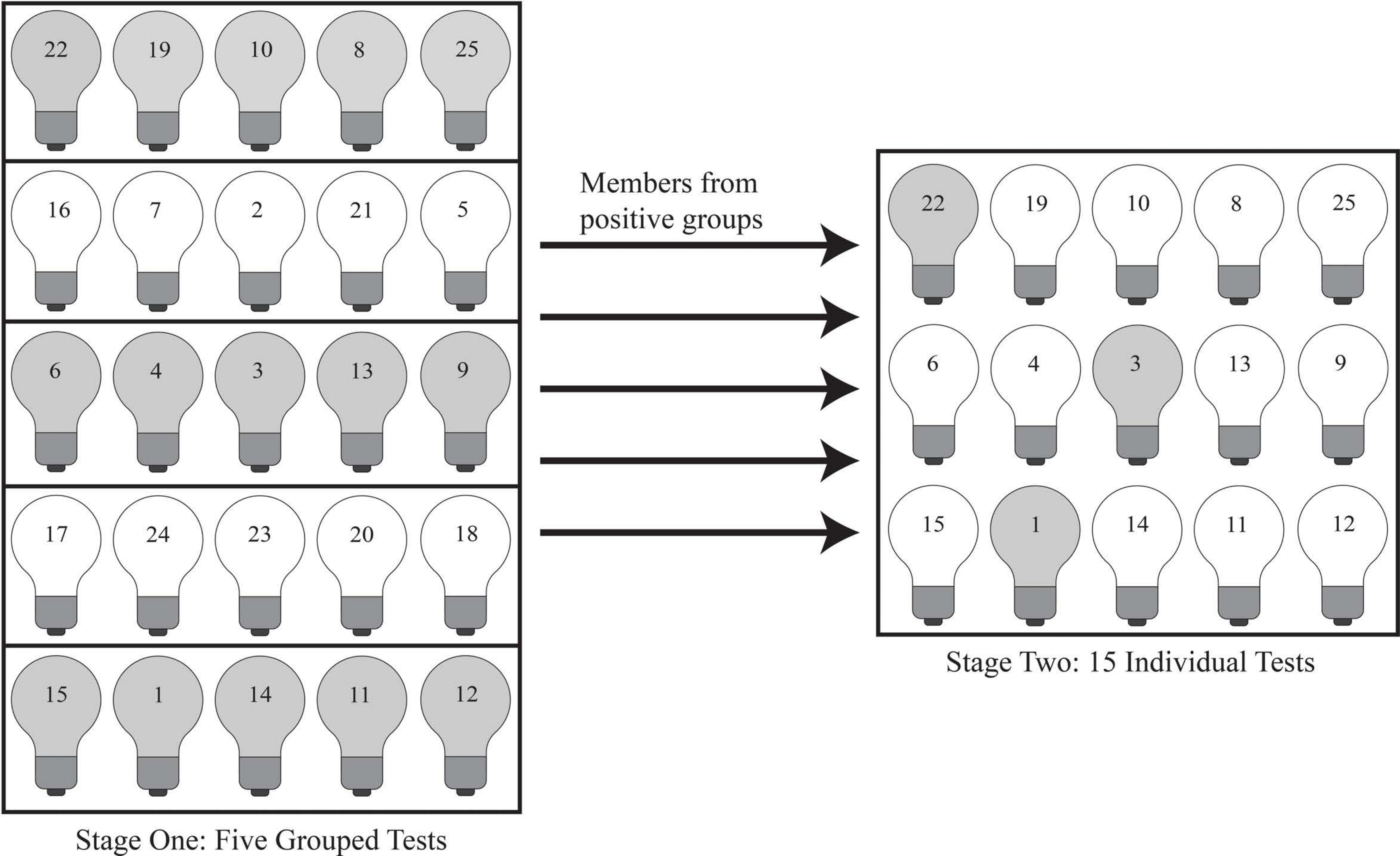}} \hfill
    \caption{Illustration of the application of the two-stage $(1,s)$-regular design algorithm (1SA) for 25 light bulbs arranged in a batch under perfect testing conditions. Light bulbs suspected to be defective by a test are colored gray. True defective light bulbs are numbered 1, 3, and 22.}
    \label{fig:dorf}
\end{figure}

The classical two-stage GT algorithm as proposed by \cite{dorfman1943}, where individual retesting occurs in the second stage is depicted in Figure \ref{fig:dorf}. Within the framework of multi-stage RSA, Dorfman’s algorithm can be classified as a two-stage $(1,s)$-regular design algorithm (1SA). \citep{aldridge2022, aldridge2022b}.
In the first stage of the two-stage 1SA, the light bulbs are grouped randomly into five groups of group size five. Thus, five group tests are conducted, revealing that only groups two and four consist of functional light bulbs. Since the remaining groups do not pass the test, it can be concluded that at least one light bulb in each of these groups is defective. To identify the exact defective bulbs, individual retesting is performed on the positive groups. The results show that the three defective light bulbs are identified after 15 individual tests. In total, the algorithm succeeds with 20 tests \citep{dorfman1943, watson1961}.

\begin{figure}[H] 
    \centering
    \makebox[\textwidth]{\includegraphics[width=0.6\textwidth]{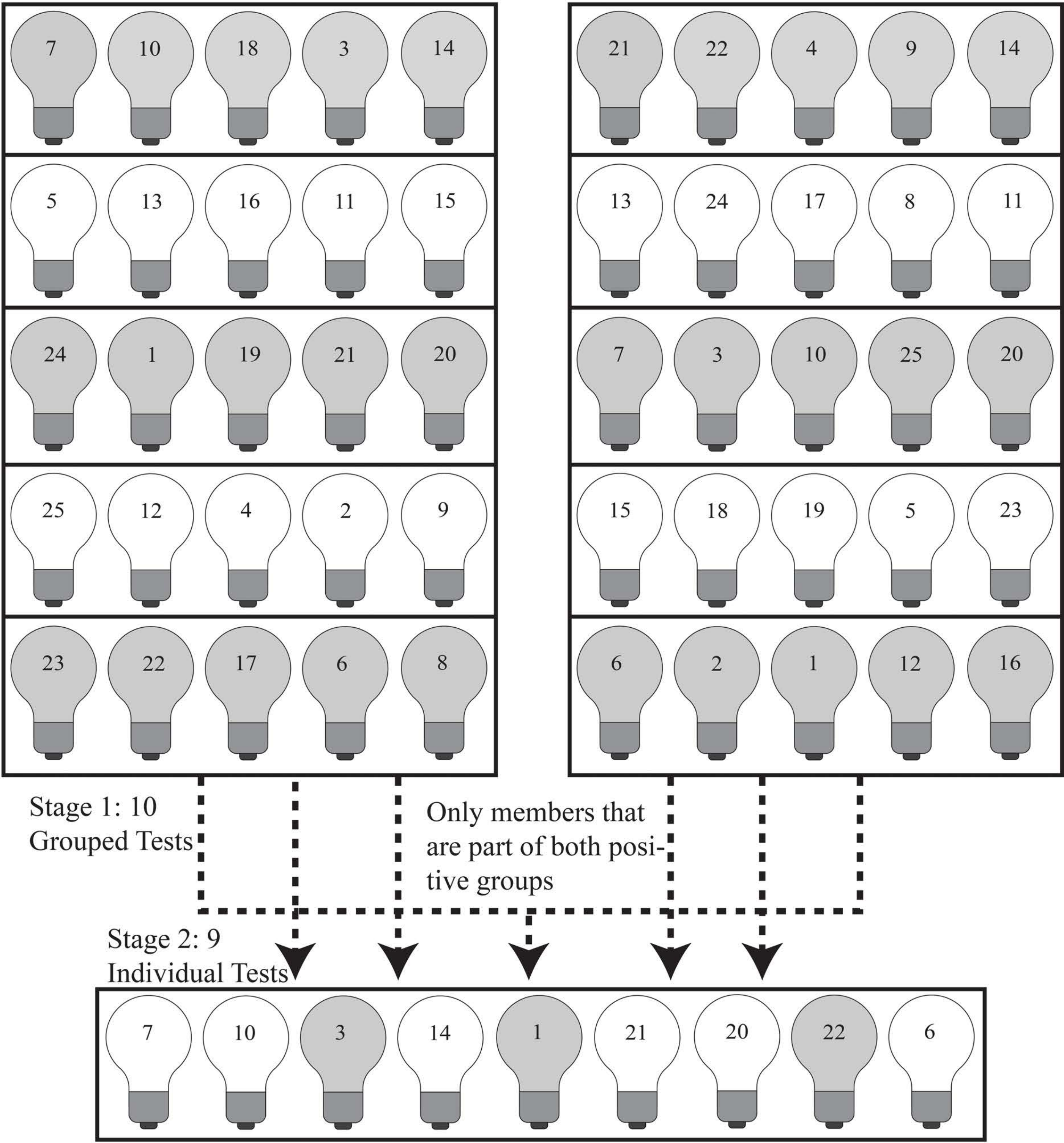}} \hfill
    \caption{Illustration of the application of the two-stage $(2,s)$-regular design algorithm (2SA) for 25 light bulbs arranged in a batch under perfect testing conditions. Light bulbs suspected to be defective by a test are colored gray. True defective light bulbs are numbered 1, 3, and 22.}
    \label{fig:dp}
\end{figure}

Figure \ref{fig:dp} illustrates the procedure of the two-stage $(2,s)$-regular design algorithm (2SA). In the first stage, the light bulbs are randomly sampled twice. Alternatively, the grid structure, that is, the "rows" and "columns" of the batch, can be utilized if parallel testing is of interest. Group tests are then conducted for each set of samples or corresponding "rows" and "columns". Particularly, the light bulbs are grouped into five groups of group size five, requiring ten group tests in total. Only light bulbs that are part of positive groups in both joint tests move to the second stage of individual retesting. As shown, groups one, three, and five test positive. The intersection of positive groups from both rounds of testing results in nine light bulbs being tested individually in the second stage. This algorithm successfully identifies all defective light bulbs with 19 tests \citep{broder2020, aldridge2022b}.

\begin{figure}[H] 
    \centering
    \makebox[\textwidth]{\includegraphics[width=0.8\textwidth]{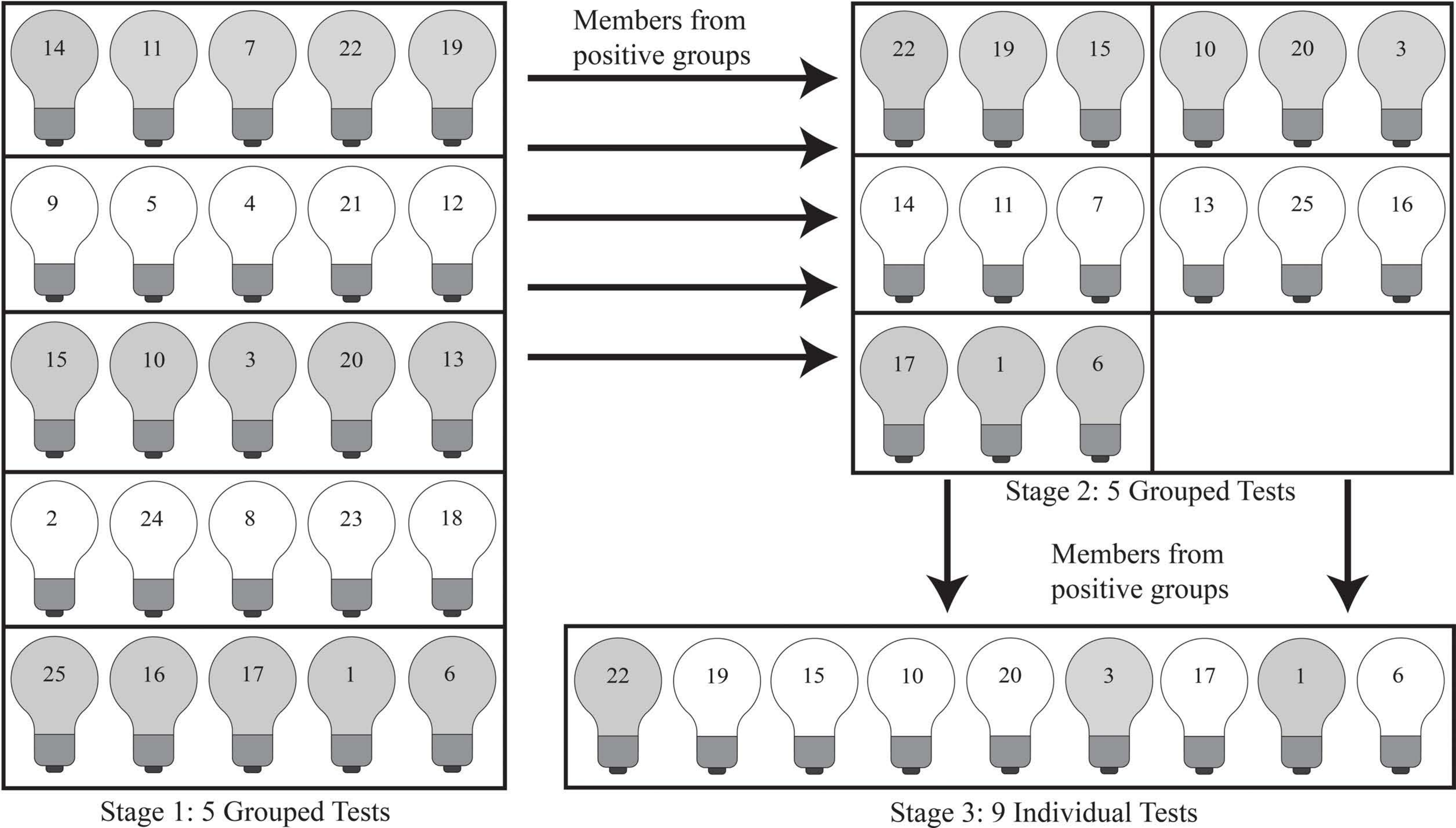}} \hfill
    \caption{Illustration of the application of the three-stage $(1,s)$-regular design algorithm (1SA) for 25 light bulbs arranged in a batch under perfect testing conditions. Light bulbs suspected to be defective by a test are colored gray. True defective light bulbs are numbered 1, 3, and 22.}
    \label{fig:three}
\end{figure}

Two-stage 1SA can easily be extended to multi-stage 1SA as noted in \cite{patel1962}. In Figure \ref{fig:three}, the three-stage 1SA algorithm is illustrated where the light bulbs in the batch are sampled randomly. Similar to two-stage 1SA, the light bulbs are divided into disjoint groups of group size five, resulting in five group tests. Three groups, namely one, three and five test positive and proceed to the second stage. In the second stage, the light bulbs from the positive groups are regrouped into smaller groups of group size three, and five additional group tests are conducted. This stage clears six light bulbs as negative, leaving nine light bulbs with an ambiguous state. These remaining nine light bulbs are then individually tested in the third stage, resulting in a total of 19 tests. In this scenario, three-stage SP performs similarly to two-stage DP in terms of the number of tests required.

\begin{figure}[H] 
    \centering
    \makebox[\textwidth]{\includegraphics[width=0.6\textwidth]{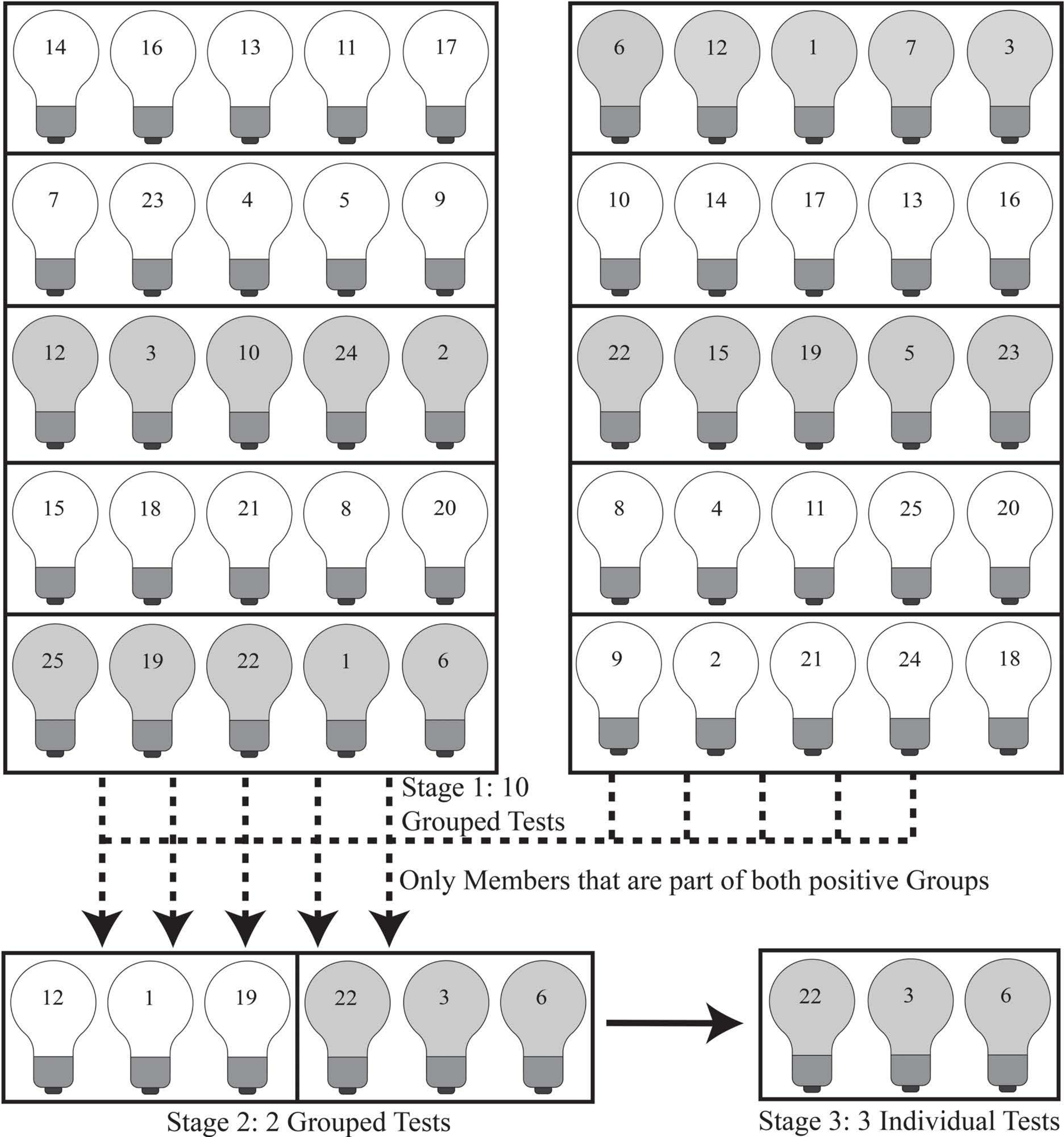}} \hfill
    \caption{Illustration of the application of the three-stage $(2,s)$-regular design algorithm (2SA) for 25 light bulbs arranged in a batch under perfect testing conditions. Light bulbs suspected to be defective by a test are colored gray. True defective light bulbs are numbered 1, 3, and 22.}
    \label{fig:threedp}
\end{figure}

Finally, a variation of the novel multi-stage RSA is presented which combines two-stage 2SA with three-stage 1SA to form three-stage 2SA. The illustrative scheme is shown in Figure \ref{fig:threedp}. In the first stage, the light bulbs in the batch are randomly sampled twice, similar to two-stage 2SA. Alternatively, the grid structure can be leveraged again. Thus, each light bulb undergoes two joint tests in which the bulbs are divided into groups of size five, requiring ten group tests overall. Light bulbs that are part of positive groups in both joint tests proceed to the second stage. In the second stage, the remaining light bulbs are grouped again into groups of size three. Two group tests are conducted where one returns negative. Consequently, three of the remaining six light bulbs can be cleared as non-defective. The final three light bulbs undergo individual testing in the third stage, allowing for the precise identification of all defective light bulbs. Three-stage 2SA successfully identifies all defective light bulbs with only 15 tests, demonstrating an improvement in efficiency compared to both three-stage 1SA and two-stage 2SA.

\section{Group Testing with Multi-stage (r,s)-regular design Algorithms} \label{sec:meth}
\subsection{Preliminaries}
In Section \ref{sec:intro}, GT is described as a general framework encompassing any algorithm concerned with identifying defective members in a population by dividing them into groups and evaluating their joint test results. However, a large body of literature effectively divides GT into two distinct categories. For instance, probabilistic GT assumes an apriori probability distribution on the defective members in the population. The primary goal in this category is to quantify the ENT required to identify all defective members. By leveraging probabilistic models, GT algorithms are then optimized with respect to the ENT. In contrast, combinatorial GT is more general and operates under a deterministic model. It does not assume any probability distribution on the population. Instead, the focus is on determining the minimal number of tests needed to guarantee the identification of all defective members, regardless of their distribution. This makes combinatorial GT applicable in scenarios where no apriori information about the distribution of defects is available \citep{du1999}. Within these both categories, GT algorithms are additionally classified as either adaptive or non-adaptive. In non-adaptive GT algorithms, all testing designs for each stage are pre-planned before any tests are conducted. This means that the entire testing algorithm is fixed from the start, and the tests are carried out simultaneously or in a predetermined sequence without taking into account the intermediate results. Adaptive GT algorithms follow a sequential procedure where the design of each stage depends on the results obtained from previous stages. This allows for dynamic adjustments based on real-time feedback, potentially reducing the number of tests needed as more information is gathered throughout the process \citep{aldrige2019, aldridge2022}.

Based on this classification scheme, multi-stage RSA can be categorized as adaptive, probabilistic GT algorithms. Besides this general classification, multi-stage RSA can be further divided into special sub-classes. For instance, there are different strategies for handling the final stage of testing in multi-stage RSA. One approach is to declare members of negative groups as non-defective and all members of positive groups as defective without any follow-up tests. This method is straightforward but may result in a higher number of false positives. Another strategy involves conducting individual follow-up tests for members who are not part of at least one negative group in $r$ joint tests. This approach balances the need to reduce false positives while minimizing the total number of tests. A third method, described in Section \ref{sec:intro} and \ref{sec:mot}, involves administering individual tests only to members that are part of all $r$ joint tests. Multi-stage RSA employing this method are often referred to in the literature as "trivial" or "conservative" \citep{aldridge2022b}. Furthermore, there are various methods for constructing the $(r,s)$-regular design in multi-stage RSA. One option is to use coding theory, where an error-correcting linear code with appropriate parameters is chosen to construct the $(r,s)$-regular design. Interested readers can refer to \cite{kautz1964} and \cite{aldrige2019} for more details on this approach. Another method involves hypercube constructions, which can be viewed as a generalization of two-stage 2SA using a grid structure. Instead of randomly sampling the population, members are arranged in a grid of rows and columns and tested in parallel. The hypercube construction extends this concept for $r > 2$. Multi-stage RSA with a hypercube design have already been investigated and practically applied. More details and applications can be found in \cite{mutesa2021} and \cite{wu2022}. The key difference between the approach in this paper and the hypercube design lies in the design structure. Instead of constructing a hypercube design, it is also possible to choose a uniformly random design where the members in a population are shuffled $r$ times. In each of these random shuffled samples, GT is conducted separately and independently. The results for subsequent stages are then obtained by combining the results from independent joint tests \citep{aldridge2022b}. Therefore, the focus in this paper is on multi-stage RSA within the framework of adaptive, probabilistic, and conservative GT, utilizing a random $(r,s)$-regular design. Given the probabilistic nature of multi-stage RSA, the aim is to quantify the ENT.

\subsection{Expected Tests in Multi-stage (r,s)-regular design Algorithms}
From the introduction to multi-stage RSA in Section \ref{sec:intro} and the illustrations in Section \ref{sec:mot}, it can be inferred that the number of tests in a GT algorithm cannot be trivially quantified. Specifically, the number of tests depends on a complex combination of factors, including the population size, the probability of a member being defective, and the group size at each stage. Furthermore, in multi-stage RSA, another layer of complexity arises due to additional joint tests at each stage which has consequences on the isolation of potentially defective members, henceforth called suspected, which state currently remains ambiguous. Particularly, multi-stage RSA can be best understood as a natural extension of two-stage RSA. In two-stage RSA, suspected members receive individual follow-up tests after the first stage of GT. Instead of applying individual follow-up tests immediately, multi-stage RSA introduces additional GT stages. In this extended approach, the suspected members are grouped again into new groups, potentially with different sizes and a different number of joint tests. This postpones individual retesting of suspected members to later stages, motivated by the potential further reduction in the number of tests achieved by clearing more members through group stages.

To formally derive the ENT $E(T)$, the current mathematical notation has to be extended and assumptions have to be imposed. Generally, the following model setup for derivation is consistent with the probabilistic GT literature \citep{aldrige2019}. Viewing the number of tests as a random variable is the standard approach in probabilistic GT literature \citep{du1999}. Thus, let the number of tests in any multi-stage RSA be denoted by $T$, which is distributed according to an unknown probability distribution $F_T$. Further, let $l \in \{1,\dots, k\}$ denote the current stage of the multi-stage RSA, where $k \geq 1$ represents the maximum number of stages without individual retesting. In each stage $l$, both the group size and the number of joint tests are allowed to vary. I denote the group size by $s_l$ and the number of joint tests by $r_l$. Individual retesting is then conducted in the final stage $k + 1$. By construction, the population in each stage $l$ consists of members drawn from the initial population size $n$, meaning $n_l$ satisfies $n_l \leq n \ \forall \ l$ with $n \in \mathbb{N}$. Assume that the state of each member, denoted by $X_i$ with $i \in \{1, \dots, n\}$, is an identically and independently distributed (i.i.d.) Bernoulli random variable with probability density function
\begin{equation*}
f_{X_i}(v,p) = \begin{cases} 
p & \text{if } v = 1, \\
q = 1 - p & \text{if } v = 0,
\end{cases}
\end{equation*}
where every member $i$ in a population is either defective with probability $P(X_i = 1) = p$ or non-defective with probability $P(X_i = 0) = q = 1 - p$. Since the population size is fixed and the Bernoulli trials are assumed to be identical and independent, the sum $\sum_{i=1}^{n} X_i$ is a Binomial distributed random variable with probability density function
\begin{equation*}
    f_{\sum_{i=1}^{n} X_i}(d,n,p) = \binom{n}{d}p^d(1-p)^{n-d},
\end{equation*}
where $d \in \mathbb{N}$ represents the number of defective members in the population.
Imposing this distributional assumption on the state of the members has implications on the apriori knowledge of defectiveness. Particularly, each member has an identical and independent probability of defectiveness which in turn leads to identical and independent apriori knowledge of defectives in the population. Specifically, the number of defective members in any population of size $n$ is fixed at $d$ defective members, defined as $d = np$. In many applications, however, the number of defective members decreases as the population size increases. If the population size can be (artificially) increased, then $d$ is constantly decreasing due to the dilution of defective members, leading to a decreasing probability of defectiveness. In the context of this paper, the number of defective members in a population is assumed to remain constant even as the population size increases. To ensure that the number of defectives is indeed proportional to the product of the population size and the probability of defectiveness, both $n$ and $p$ must remain constant. Specifically, $p$ is assumed to be constant as $n \to \infty$ which indicates the linear regime. On the same note, it is assumed that the population to be tested is sufficiently large. Due to the underlying combinatorial considerations, a small $n$ immediately results in fewer possible ways to arrange the members into groups. Consequently, the number of possible test outcomes decreases, causing the average number of tests to fluctuate rather than converge to the ENT. For mathematical convenience the so-called zero-error paradigm or perfect testing condition is imposed which ensures that all defective members can be identified with complete certainty, eliminating concerns of sensitivity and specificity in test results. In most cases, closed-form expressions can only be obtained asymptotically, where asymptotic equivalence $\stackrel{a}{\sim}$ is defined as $\frac{a(n)}{b(n)} \to 1$ as $n \to \infty$ \citep{aldrige2019, aldridge2022, aldridge2022b}.

Based on these assumptions, lemmata can be derived, which are useful for the derivation of the central theorems concerning the ENT. Particularly, the quite general Lemma \ref{lm:bin} is helpful in deriving the probabilities occurring in multi-stage RSA. A proof is shown in Appendix \ref{app:lm_bin}.

\begin{lemma} \label{lm:bin}
In any stage $l$, let a population of size $n_l$ be much larger than a group size $s_l$. Then, for any $s_l = o(\sqrt{n_l})$, following relationship holds asymptotically
\begin{equation*}
    \binom{n_l}{s_l} \stackrel{a}{\sim} \frac{n_{l}^{s_l}}{s_l!}.
\end{equation*}
\end{lemma}

In multi-stage 1SA, the population is divided into groups once in the first stage. For subsequent stages, the results from the first round of group tests are necessary to make decisions about further testing. Therefore, it is essential to investigate the label of the groups after testing. Depending on the label of the groups, members can be directly declared non-defective due to the assumption of perfect testing conditions. Likewise, members that are subject to further testing can be isolated. Furthermore, the number of tests in any stage with grouping depends on the number of groups in that exact stage. Lemma \ref{lm:groups} quantifies the probability of the group labels. A proof is given in Appendix \ref{app:lm_groups}.

\begin{lemma} \label{lm:groups}
In any stage $l$, let a population of size $n_l$ be divided into disjoint groups $G_l$, where each group $g_l \in G_l$ is of equal size $s_l$. Furthermore, let each member $i \in g_l$ be identically and independently non-defective with probability $P(X_i = 0) = q = 1-p$. Then the probability that group $g_l$ is negative, meaning all members are non-defective, is $P(\text{$g_l$ is negative}) = q^{s_l} = (1-p)^{s_l}$.
\end{lemma}

Furthermore, Lemma \ref{lm:groups} implies that the label of the groups can be interpreted as an i.i.d. Bernoulli random variable with probability density function given by 
\begin{equation*}
f_{g_l}\left(u,p^{s_l}\right) = \begin{cases} 
p^{s_l} & \text{if } u = 1, \\
q^{s_l} = 1 - p^{s_l} & \text{if } u = 0,
\end{cases}
\end{equation*}
meaning that a group in stage $l$ is either positive $P(\text{$g_l$ is positive}) = P(g_l = 1) = p^{s_l}$ or negative $P(\text{$g_l$ is negative}) = P(g_l = 0) = q^{s_l}$. For multi-stage 1SA, quantifying the number of positive and negative groups is sufficient to make decisions about the further testing procedure. However, in multi-stage RSA, the label of the groups in each independent joint test $r_l > 1$ carries only partial information for decision-making about further testing. Specifically, the focus is not strictly on the members of positive groups in a joint test, but rather on the members that are part of positive groups in $r_l$ joint tests. Thus, the results of an independent joint test are not sufficient to declare members as either defective or non-defective. Therefore, the state of each member in each stage $l$ is subject to ambiguity. Thus, let $Y^{(l)}_i$ be an i.i.d. Bernoulli random variable describing the ambiguity of the state of each member in stage $l$. Then a member is termed suspected if its state is ambiguous. Since in multi-stage RSA only suspected members move to the subsequent stages of additional testing, the probability of ambiguity of the state must be quantified in order to compute the ENT.  

\begin{lemma} \label{lm:sus}
In any stage $l$, let a population $n_l$ be divided into equal sized $s_l$ disjoint groups $G_l$ for a total of $r_l$ times at random. Furthermore, let each member $i$ be identically and independently non-defective with probability $P(X_i = 0) = q = 1-p$. Then a member is called suspected, meaning the state is currently ambiguous, with probability $P\left(Y^{(l)}_i = 1\right) = p + q(1 - q^{s_{l}-1})^{r_{l}}$.
\end{lemma}
Lemma \ref{lm:sus} quantifies the probability of the ambiguity of the state and implies that the probability density function is given by
\begin{equation*}
f_{Y^{(l)}_i}(m,p + q(1 - q^{s_{l}-1})^{r_{l}}) = \begin{cases} 
p + q(1 - q^{s_{l}-1})^{r_{l}} & \text{if } m = 1, \\
1 - (p + q(1 - q^{s_{l}-1})^{r_{l}}) & \text{if } m = 0,
\end{cases}
\end{equation*}
which is helpful in capturing the probability of propagated ambiguity throughout multiple stages of multi-stage RSA. Thus, the state of a member is considered ambiguous in any stage $l$ after the results of the joint test return, if the member is either truly defective or non-defective but part of positive groups in $r_l$ joint tests. A proof is given in Appendix \ref{app:lm_sus}. Combining the assumptions, lemmata, and the general description of multi-stage RSA, the central theorem of this paper quantifying the ENT is presented in Theorem \ref{theorem:cent}.

\begin{theorem} \label{theorem:cent}
If the population size is sufficiently large $n \to \infty$, GT with adaptive, probabilistic and conservative multi-stage RSA can be completed in
    \begin{equation*}
        E(T) \stackrel{a}{\sim} n \left( \frac{r_1}{s_1} + \sum_{l=2}^{k} \frac{r_l}{s_l} \left(p + q(1-q^{s_{l-1}-1})^{r_{l-1}}\right) + p + q(1-q^{s_k-1})^{r_k} \right).
    \end{equation*}
tests on average.
\end{theorem}

Given the expression in Theorem \ref{theorem:cent}, it can be concluded that the ENT depends on the size of the population $n$, the probability of being defective $p$, the constant members per test $s_l$, and the constant number of joint tests per member $r_l$. Typically, it is assumed that $n$ is known, and $p$, and thus $q$, are either also known or can be estimated based on previously available information. In contrast, this implies that $s_l$ and $r_l$ are not given exogenously and can, in principle, be freely chosen. In practice, however, $s_l$ and $r_l$ are restricted by the practitioner through the choice of the maximum number of stages $k$. Additionally, it is often not desirable to choose $s_l$ and $r_l$ freely because this could result in suboptimal ENT. Instead, the following optimization problem can be solved to ensure that $s_l$ and $r_l$ are chosen such that the ENT is minimized
\begin{subequations}
\begin{alignat*}{2}
&\!\min_{s_l,r_l}        &\qquad& E(T)\\
&\text{subject to} &      & s_l \geq 1,\\
&                  &      & r_l \geq 1.
\end{alignat*}
\end{subequations}
Independent of the choice of $k$, it is evident that attempting to derive the partial derivatives of $E(T)$ with respect to $s_l$ and $r_l$ and to solve the first-order condition results in an expression that is not in closed form. Thus, numerical optimization techniques have to be employed to quantify the values at which the ENT is minimal. However, the problem becomes increasingly complex depending on the choice of $k$. For example, choosing $k = 3$, which is equivalent to four-stage RSA, already yields a six-dimensional nonlinear optimization problem. In this case, the values of the three group sizes and joint tests have to be optimized simultaneously. To ease the computational burden and make optimization feasible for a high number of stages, restrictions on the $r_l$ can be imposed. Naturally, the practitioner will reserve the option to decide how many joint tests can be performed based on the available capacity. Therefore, one option is to set all $r_l$ beforehand, reducing the number of parameters in the optimization by half. Another option would be to set the $r_l$ to a common $r$ in all stages, effectively condensing the number of optimization parameters for the joint tests to one. Thus, through numerical optimization with certain user-specific tweaks, the $s_l$ and $r_l$ can be chosen such that the ENT is minimal.

\subsection{The Family of Multi-stage (r,s)-regular design Algorithms}
In the discussion of the optimization of the ENT, it becomes clear that the practitioner plays a crucial role in the choice of the particular multi-stage RSA. Allowing the parameters to vary freely provides maximal freedom but severely reduces their practical applicability. Recall that two-stage RSA are accompanied by severe limitations, making their practical applicability for higher stages infeasible. In fact, the benefits in terms of ENT quickly diminish because in each stage potentially multiple $r_l$ joint tests are necessary. Therefore, the focus in practice is rather on important special cases that can be recovered from Theorem \ref{theorem:cent}. These special cases have the central benefit of being less complex and thus much more applicable in real-world scenarios. Moreover, some of them have been extensively studied in GT literature, where Theorem \ref{theorem:cent} can be viewed as a unifying expression for the ENT.

Setting the maximum number of stages to $k = 2$ recovers two-stage RSA discussed in \cite{broder2020} and \cite{aldridge2022b}. The expression for the ENT is then given by
\begin{equation*}
    E(T) \stackrel{a}{\sim} n \left( \frac{r}{s} + \left(p + q(1-q^{s-1})^{r}\right)  \right).
\end{equation*}
Two-stage 2SA is a special case of two-stage RSA, which can be recovered by setting $r = 2$. Consequently, the two-stage 1SA algorithm discussed in \cite{dorfman1943} is obtained by setting $r = 1$.

An expression for the ENT in multi-stage 1SA can be obtained by considering an arbitrary number of stages $k>2$ while setting the number of joint tests in each stage $l$ to $r_1 = r_l = r_k= 1$. The ENT can then be written as 
\begin{equation*}
        E(T) = n \left( \frac{1}{s_1} + \sum_{l=2}^{k} \frac{1}{s_l} \left(1-q^{s_{l-1}}\right) + (1-q^{s_k}) \right).
\end{equation*}
This expression is identical to the derivation found in \cite{patel1962}. Note further that equality holds here due to the results of Lemma \ref{lm:groups}. It is highly unlikely that algorithms with a large $k$ are utilized in practice since they are not very applicable in real-world settings due to the potential long duration of the GT algorithms. Therefore, there is usually an upper bound on $k$ chosen by practitioners. A common choice is usually $k = 4$ \citep{patel1962}.

All the presented special case GT algorithms have been extensively investigated in the literature and practically applied in the real world \citep{mutesa2021, aldridge2022b}. The novel Theorem \ref{theorem:cent} for the ENT allows for the utilization of a flexible set of multi-stage RSA. Thus, a wide range of $r_l$ and $s_l$ combinations across multiple stages can be utilized to further reduce the ENT. It is important to point out that the choice of especially $r_l$ and $k$ should always be made carefully, as there is no theoretical restriction on the number of stages or joint tests. However, it is very likely that an improper choice might lead to multi-stage RSA where the ENT is much larger than that of established GT algorithms. A common misleading example is choosing $k = 3$ but setting every $r_l$ to a very large value. Since the number of tests due to the joint tests of $r_l$ quickly outweighs any advantages gained by clearing more members in the stages, the ENT will be much larger than for simpler GT algorithms. Nevertheless, the flexible family of multi-stage RSA still allows for finding GT algorithms that combine practical applicability with a further reduction of ENT compared to established GT algorithms. For instance, instead of utilizing a $(r,s)$-regular design in every stage of a multi-stage GT algorithm, a combination of a $(r,s)$-regular design with $(1,s)$-regular design can be applied. Then in the first stage the number of joint tests can be set to $r_1 = r$ and in subsequent stages to $r_l = r_k = 1$. The expression for the ENT is then given by
\begin{equation*}
    E(T) \stackrel{a}{\sim} n \left( \frac{r}{s_1} + \frac{1}{s_2} \left(p + q(1-q^{s_{1}-1})^{r}\right) + \sum_{l=3}^{k} \frac{1}{s_l} \left(1-q^{s_{l-1}}\right) + (1-q^{s_k}) \right). 
\end{equation*}
However, as discussed before, the GT algorithm can be further simplified by setting $r=2$ and thus retaining the general $(2,s)$-regular design structure. Then the resulting GT algorithm is another special case of the multi-stage 2SA algorithm.

\section{Computational Simulations}  \label{sec:num}
\subsection{Performance Criteria and Study Design}
Theorem \ref{theorem:cent} shows asymptotic equivalence for the ENT. However, in all real-world scenarios the underlying population is finite. To investigate the behavior of the ENT in finite populations, the mathematical derivation are supported by rigorous simulation studies. To this end, I introduce the general study design as well as performance criteria for the comparison of GT algorithms. Subsequently, simulation studies and performance evaluations are conducted based on the introduced criteria via computational simulations. By considering common performance criteria, the novel multi-stage RSA are compared to the already established GT algorithms. Typically, the standard performance criterion in GT literature is the expected tests per member (ETM), denoted by
\begin{equation*}
   \text{ETM} = \frac{E(T)}{n}.
\end{equation*}
Generally, a lower ETM indicates better performance of a GT algorithm and is thus always preferred. It can also be directly observed, that for the case of individual testing the expression simplifies to $\frac{E(T)}{n} = \frac{n}{n} = 1$. Therefore, the ETM effectively serves as a criterion for comparing GT algorithms to the individual testing algorithm as the upper limit. Additionally, it is possible to establish a counting bound, which serves as the lower bound for any GT algorithm. In information theory, the counting bound is expressed using the binary entropy function
\begin{equation*}
    H(p) = -p \log_2(p) - (1-p) \log_2(1-p).
\end{equation*}
The following inequality then holds
\begin{equation*}
    E(T) \geq H(p)n.
\end{equation*}
The performance criterion based on the binary entropy function is called the rate and is defined as 
\begin{equation*}
    \text{Rate} = \frac{H(p)n}{E(T)},
\end{equation*}
which measures the closeness of a GT algorithm to the counting bound. It can be interpreted as the information gain per expected test. Higher values are always preferred \citep{aldrige2019,aldrige2019b}.

Regarding the study design, I compare six different GT algorithms: two-stage, three-stage, and four-stage 1SA, 2SA, and RSA. However, while three- and four-stage 2SA typically involve joint tests at every stage, I focus on GT algorithms where joint tests are only applied in the first stage. To clearly differentiate between these algorithms, I adjust the classical labels of the two-stage RSA. Specifically, two-stage 1SA, 2SA, and RSA are abbreviated as SP-Two, DP-Two, and RP-Two; three-stage 1SA and 2SA as SP-Three and DP-Three; and four-stage 1SA and 2SA as SP-Four and DP-Four. Here, SP, DP, and RP stand for single pooling, double pooling, and $r$-pooling which indicate the cardinality of joint tests that are conducted in the first stage $r_1 = r$ of the GT algorithms. Thus, stages beyond the first are always set to $r_l = r_k = 1$ \citep{aldridge2022b}. All GT algorithms are selected for their practicality in real-world application settings. Furthermore, I have already provided an introduction to these special cases of multi-stage RSA in Section \ref{sec:mot} and \ref{sec:meth}. As a benchmark GT algorithm for comparison, I choose two-stage RSA since it approaches the counting bound but has certain drawbacks when applied in real-world applications \citep{aldridge2022b}. In the presentations of the results, the population sizes are fixed at $n = 100$ and $n = 1000$. The probability $p$ of a member being defective is given by $p \in \{0, 0.001, \dots, 0.35\}$ with a step length of $0.001$. For each $p$, the ENT is then optimized with respect to unknown group size $s_l$ and joint tests $r$. Afterward, the testing procedure is simulated with optimal group size for $m_{\text{val}} = 100$ times at each $p$. I will then compare the ETM with the average number of tests per member (ATM)
\begin{equation*}
    ATM = \frac{\bar{T}}{n},
\end{equation*}
where the $\bar{T}$ is defined as
\begin{equation*}
    \bar{T} = \frac{1}{m_{\text{val}}} \sum_{i = 1}^{m_{\text{val}}} T_i.
\end{equation*}
Furthermore, the range is provided as a measure of uncertainty given by
\begin{equation*}
    R = T_{\text{max}} - T_{\text{min}}.
\end{equation*}
To capture potential deviations between the ETM and ATM occurring due to the simulation of the GT algorithms with finite $n$, an approximation error is additionally introduced. Specifically, the mean absolute percentage error (MAPE) is given by
\begin{equation*}
    \text{MAPE} = \frac{1}{\left|p\right|} \sum_{t=1}^{\left|p\right|} \left| \frac{E(T) - \bar{T}}{E(T)} \right| \cdot 100,
\end{equation*}
where $\left|p\right|$ is the cardinality of $p$ and $\left| \frac{E(T) - \bar{T}}{E(T)} \right|$ represents the absolute percentage error. The MAPE is presented in probability intervals. The cuts are based on the feasibility of the GT algorithms. Particularly, the intervals are given by $p < 0.077$ where four-stage 1SA and 2SA are still feasible, $p \in (0.077, 0.182)$ where three-stage 1SA and 2SA are still feasible and $p > 0.182$ where only two-stage 1SA and 2SA remain. All simulations are performed in the programming language \textbf{R} \citep{R2025}. Optimization is additionally conducted via the \textbf{optimx} package \citep{nash2011, nash2014}. The presented graphics are created with the \textbf{tidyverse} packages \citep{tidyvere2019}. The code for the reproducibility of all results is freely available in the GitHub repository \url{https://github.com/micbalz/MultiGT}.

\subsection{Results}
\begin{figure}[!htpb] 
    \centering
    \includegraphics[width = \textwidth]{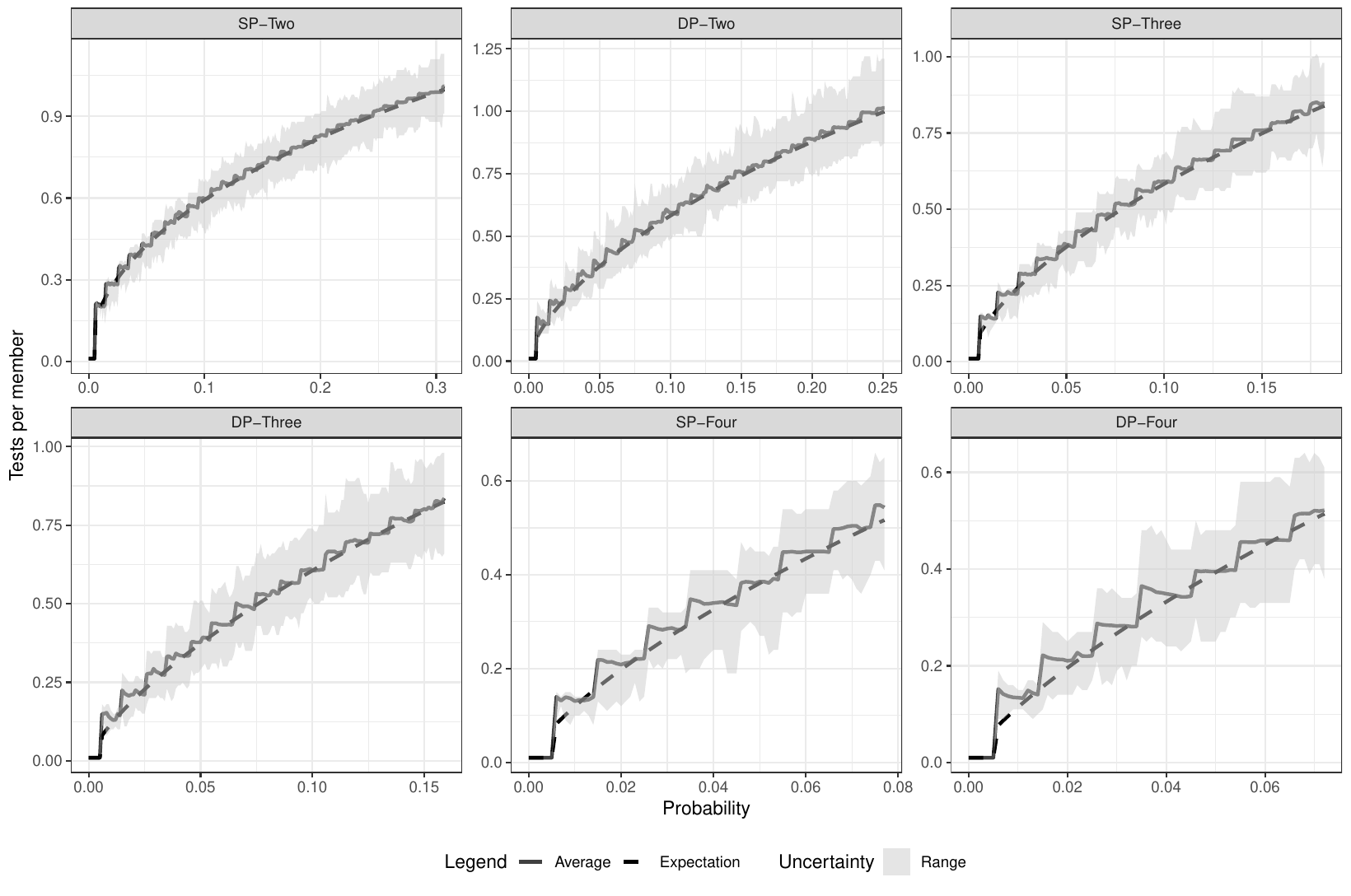} \hfill
    \caption{Comparison of expected number of tests (ETM) pictured as black dashed lines and average number of tests (ATM) pictured as gray straight lines for admissible probabilities $p$ and small population size $n = 100$.}
    \label{fig:sim1}
\end{figure}

\begin{figure}[!htpb] 
    \centering
    \includegraphics[width = \textwidth]{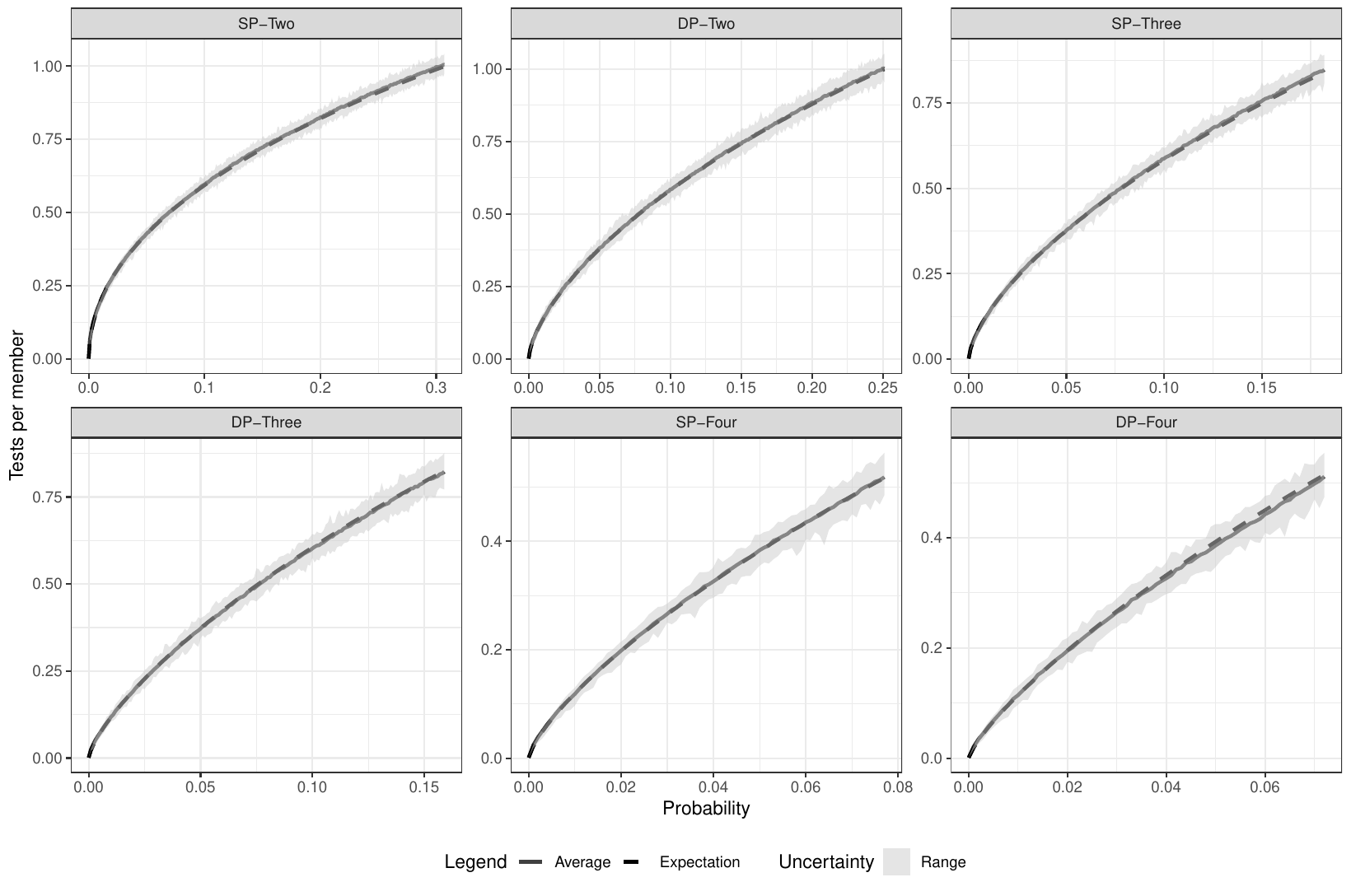} \hfill
    \caption{Comparison of expected number of tests (ETM) pictured as black dashed lines and average number of tests (ATM) pictured as gray straight lines for admissible probabilities $p$ and small population size $n = 1000$.}
    \label{fig:sim2}
\end{figure}

I start the discussion of the results by evaluating the ETM, ATM, and MAPE. To this end, the ETM and ATM are compared for all admissible values of $p$ in the GT algorithms. In Figure \ref{fig:sim1}, the results of the simulations for $n = 100$ are shown.
The results indicate that for small but practically realistic values of $n$, the ATM represented by the solid line exhibits a jumping behavior across different values of $p$. The ETM is represented by a dashed line. This behavior can be attributed to the limited number of possible combinations for assigning members to groups. Depending on $p$, optimal group sizes tend to be quite large sometimes leading to an overestimation of the ATM. For instance, this can lead to all groups being either positive or negative with very few outcomes in between at each iteration of the simulation. Thus, the defective members may all be placed in separate groups resulting in only positive group test results. Since groups are never cleared, the subsequent stages involves GT with the complete population. Nevertheless, the uncertainty bands representing the range show that the simulated ATM almost always falls within the range of the ETM. Similarly, the results for a larger population size $n = 1000$ are shown in Figure \ref{fig:sim2}.
In this case, with $n = 1000$, the results show smooth curves for both the ETM and ATM across all considered GT algorithms. Furthermore, the ATM almost perfectly aligns with the ETM with very little deviation observable. For high values of $p$, the SP-Three algorithm tends to slightly overestimate, while the DP-Four algorithm tends to slightly underestimate the ETM. Additionally, the previous jumps observed across changing probabilities are no longer present. Finally, uncertainty is low across all GT algorithms indicating proper functionality and a tendency toward the mean. While GT for high populations sizes is highly unlikely to be applied due to restrictions in practicality, it is nevertheless very useful for providing additional validation of the mathematical derivations of the ENT.

Lastly, I report the MAPE values for both simulation settings across different probability intervals. The results are presented in Table \ref{tab:mape}.
\begin{table}[!htpb]
\centering
\caption{\label{tab:mape}MAPE results for the expected number of tests (ENT) with very small $n = 100$ and finite $n = 1000$ population sizes for admissible probability intervals.}
\scriptsize
\setlength{\tabcolsep}{2pt} 
\begin{tabular}{l*{6}{c}}
\toprule
\multirow{2}{*}{Algorithm} & \multicolumn{3}{c}{$n = 100$} & \multicolumn{3}{c}{$n = 1000$} \\
\cmidrule(lr){2-4} \cmidrule(lr){5-7}
& $p < 0.077$ & $p \in (0.077, 0.182)$ & $p > 0.182$ & $p < 0.077$ & $p \in (0.077, 0.182)$ & $p > 0.182$ \\
\midrule
SP-Two & 4.90\% & 1.48\% & 0.977\% & 0.434\% & 0.565\% & 0.565\% \\
DP-Two & 8.34\% & 1.96\% & 1.36\% & 0.656\% & 0.288\% & 0.422\% \\
SP-Three & 6.93\% & 1.85\% & - & 0.549\% & 0.820\% & - \\
DP-Three & 9.14\% & 1.82\% & - & 0.826\% & 0.491\% & -\\
SP-Four & 7.94\% & - & - & 0.662\% & - & -\\
DP-Four & 9.45\% & - & - & 1.43\% & - & - \\
\bottomrule
\end{tabular}
\end{table}
As expected, it can be observed that for each consecutive interval, the proper functionality of the GT algorithms with a higher number of stages diminishes. Furthermore, as the number of joint tests increases, the MAPE also increases as suggested in finite samples by Theorem \ref{theorem:cent}. For $p < 0.077$, the MAPE is reported for every considered algorithm. In the interval $p \in (0.077, 0.182)$, four-stage SP and DP no longer have a MAPE value because at these probabilities, members cannot be efficiently divided into groups. In such cases, all groups in any stage are always positive leading to ATM values greater than one. Similar results are observed for three-stage SP and DP if $p > 0.182$. Comparing both scenarios, it can be concluded that the MAPE in any interval is consistently higher for $n = 100$. The highest values are reached for the interval $p < 0.077$, with MAPE values approaching 10\%. In the other intervals, the MAPE decreases to around 1\%. For $n = 1000$, the MAPE remains around 1\% across all probability intervals. SP algorithms exhibit slightly lower MAPE values than their DP counterparts for any number of stages. The slight overestimation in three-stage DP and underestimation in four-stage DP of the ATM can thus also be observed in the MAPE.

\begin{figure}[!htpb] 
    \centering
    \includegraphics[width = \textwidth]{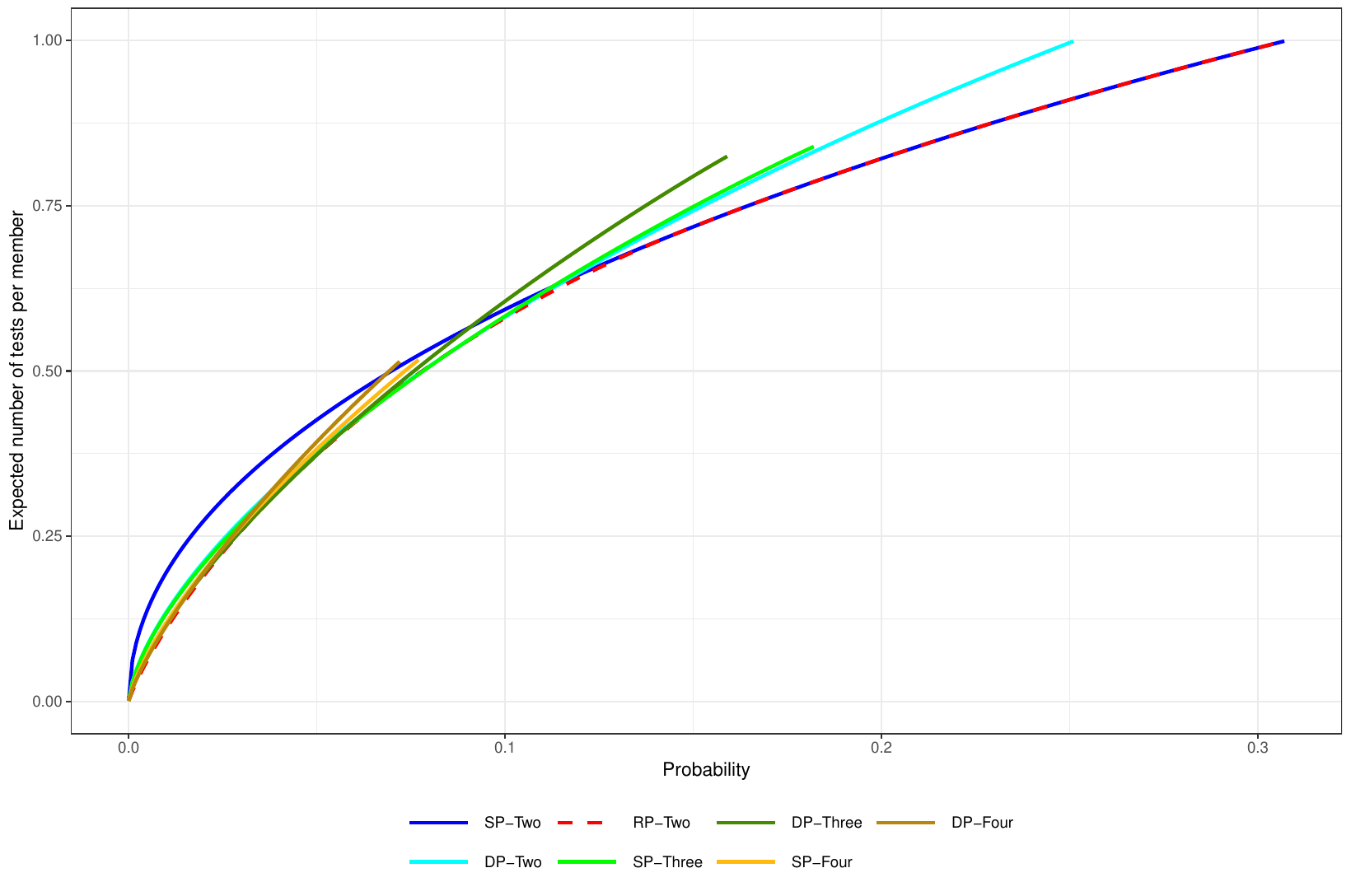} \hfill
    \caption{Performance evaluation via the expected number of tests (ETM) for eligible probabilities $p$. Each line represents the progress of the ETM in each algorithm for the complete probability range.}
    \label{fig:per}
\end{figure}

Finally, the performance of the GT algorithms is compared with a focus on evaluating the ETM across the entire eligible probability range. Since all algorithms demonstrate proper functionality for $p < 0.077$, as indicated by the MAPE analysis, more detailed insights are provided through the rate within this probability range. The ETM results are presented in Figure \ref{fig:per}.
The dashed line represents the benchmark two-stage RSA which approaches the optimal counting bound for low probability values. The goal of this study is to demonstrate that there are GT algorithms capable of achieving performance close to that of two-stage RSA without an increase in complexity. The results show that for any probability $p$, the special cases of multi-stage RSA either match or closely approximate the ETM of two-stage RSA. Additionally, the cutoff points described in the MAPE analysis are visually apparent. However, the algorithms tend to cluster around two-stage RSA across the entire probability range, making it challenging to draw distinct conclusions based solely on the ETM. Nonetheless, some patterns are observable. The two-stage 1SA exhibits the highest ETM for $p < 0.1$ but quickly becomes the algorithm with the lowest ETM as $p$ increases. This behavior is consistent with findings in the GT literature, where algorithms with a higher number of stages or greater constant tests show significant advantages in low-probability scenarios \citep{patel1962}. As the probability of defectiveness increases, these advantages diminish rapidly, leaving two-stage DP as the most viable option. For $p > 0.14$, two-stage DP becomes identical to two-stage SP indicating that using more stages or a higher number of joint tests does not reduce the ETM beyond this probability threshold.

\begin{figure}[!htpb] 
    \centering
    \includegraphics[width = \textwidth]{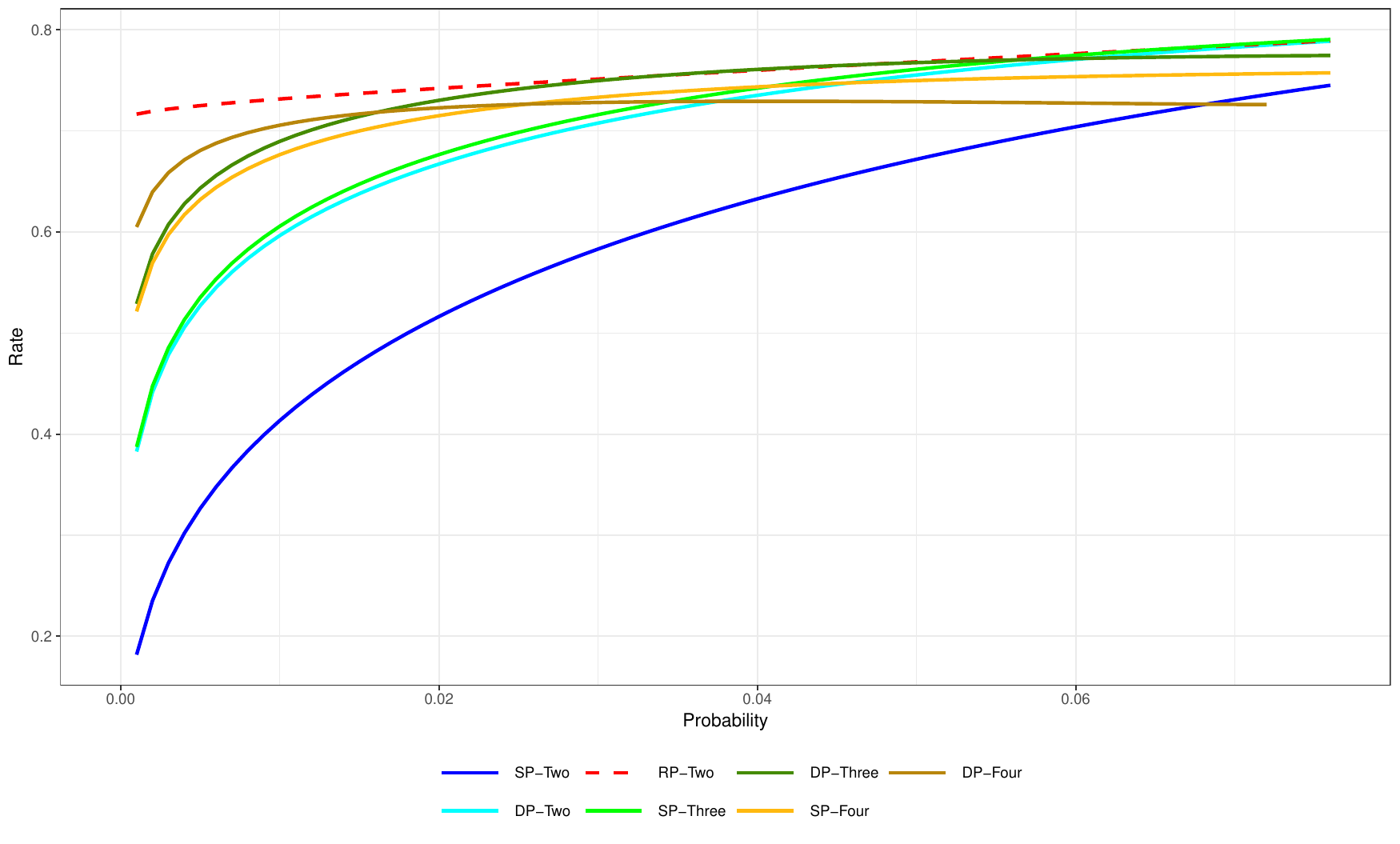} \hfill
    \caption{Performance evaluation via the rate for probabilities $p < 0.077$. Each line represents the progress of the ETM in each algorithm for the complete probability range.}
    \label{fig:per2}
\end{figure}

To provide more detailed insights into performance for $p < 0.077$, the progression of the rate within this probability range is depicted in Figure \ref{fig:per2}.
In this context, the interpretation is the inverse of the interpretation of the ETM. Consistent with earlier observations, two-stage SP has the poorest performance in terms of the rate within this probability range. A much better alternative is the established two-stage 2SP which shows significant improvement across all probabilities within this range. As suggested in the GT literature, two-stage RSA remain the best algorithms in terms of the rate. However, when comparing the established two-stage DP with the three-stage and four-stage SP, it becomes evident that these multi-stage GT algorithms outperform two-stage DP  within this probability range. This leads to the conclusion that established multi-stage SP are a viable alternative to the two-stage DP. The improvement of multi-stage SP over the two-stage SP is well-documented in GT literature \citep{patel1962}. Interestingly, the novel three-stage DP and four-stage DP outperform their SP counterparts for probabilities $p < 0.06$ by a small margin. Additionally, they achieve performance much closer to two-stage RSA without increasing practical complexity. These results suggest that the limitations of two-stage RSA can be mitigated across all probability ranges. For low probabilities $p < 0.06$, the novel three-stage DP and four-stage DP are thus viable alternatives. Therefore, by leveraging a combination of novel and established GT algorithms, the advantages of a lower ETM can be achieved without the impracticality associated with the two-stage RSA.

\section{Conclusion} \label{sec:conc}
In this paper, I propose novel multi-stage RSA for GT. Although the recently introduced two-stage RSA can significantly reduce the ETM and approach the optimal counting bound, they have certain limitations that hinder their practicality in real-world applications. Specifically, the potential high number of joint tests makes them unsuitable due to limited resources and planning constraints. Additionally, parallel testing often requires a large number of workers, which may not be readily available. To address these challenges, multi-stage DP offer a feasible alternative. This study provides a schematic illustration of these GT algorithms, with an application in quality control within manufacturing and engineering contexts. The results suggest that a combination of the novel three- and four-stage SP and DP can be effectively utilized depending on the probability for defectiveness. These GT algorithms achieve similar results in ETM while avoiding the drawbacks associated with two-stage RSA.

The key findings and main contributions of this paper are: (a) The expressions for the ENT of various special cases of GT algorithm previously introduced in the literature can be unified under a single Theorem \ref{theorem:cent} in the framework of multi-stage RSA. I additionally confirm that this unification allows for the recovery of all special cases studied in prior works such as \cite{dorfman1943, li1962, patel1962, broder2020, aldridge2022b}. (b) Since the equivalence in the expression in Theorem \ref{theorem:cent} holds only asymptotically, the mathematical proofs are supported by rigorous simulation studies for finite population sizes. These computational simulations demonstrate that for small $n = 1000$, the MAPE is approximately 1\% across all probabilities. Although the MAPE increases as the population size decreases to $n = 100$, it remains within reasonable limits, ensuring the feasibility of the GT algorithms. (c) In the evaluation of the performance, novel three- and four-stage DP algorithms outperform established multi-stage 1SA if the probability $p < 0.06$ is low. Additionally, the study compares and shows that multi-stage 1SA generally outperform the two-stage 2SA, thereby closing an important gap in current GT literature. For $p > 0.14$, two-stage RSA coincides with two-stage 1SA, making it the best algorithm in terms of ETM and efficiency until $p = 0.35$, beyond which individual testing becomes the only feasible GT algorithm. (d) This study also introduces the potential application of novel GT algorithms in quality control, using the light bulb testing problem as an illustrative example in Section \ref{sec:mot}. (e) The paper introduces a novel performance criterion based on the expected duration for members. A new theorem capturing this duration is introduced, with results supported by further simulation studies. Detailed derivations, proofs, and results can be found in Appendices \ref{sec:app}, \ref{sec:app2} and \ref{app:theorem:dur}.

Nevertheless, the methodology presented in this paper has several drawbacks, primarily due to the strict assumptions made during the derivation of the GT algorithms. Notably, the assumption of perfect testing conditions could be relaxed by introducing additional parameters to account for sensitivity and specificity. While this would complicate the mathematical derivations, it could provide valuable insights for applications where testing errors are common. Another limitation is the assumption of identical and independent apriori knowledge for defectiveness, which is often unrealistic in many real-world scenarios. Thus, incorporating the dynamics between members of a population into the GT algorithms could be a valuable extension to address this issue. Additionally, while the paper suggests using the new GT algorithms in quality control, it lacks a tangible application with real-world data to evaluate performance. As a result, the findings remain purely theoretical. I leave these questions open for further research, particularly encouraging practitioners to apply novel multi-stage RSA to reduce (economic) cost in quality control in manufacturing and industrial engineering.

\backmatter

\begin{appendices}
\section{Expected Duration in Multi-stage (r,s)-regular design Algorithms} \label{sec:app}
In various real-world applications, the ENT provides only partial information in the evaluation of the performance of GT algorithms. For instance, in epidemiology, testing is often subject to resource constraints with time being a particularly valuable component. Members involved in GT must remain in quarantine until they are labeled as non-defective. Algorithms that result in longer quarantine durations are less likely to be practical or relevant. Unfortunately, the ENT does not capture any information regarding the expected duration that members remain in the testing procedure. To address situations where time is a critical resource, I propose a novel performance criterion that incorporates the temporal aspect of the GT algorithms. This allows for a more comprehensive evaluation of GT performance, especially in scenarios where time efficiency is paramount. As a first step, the time that members spend in multi-stage RSA has to be quantified. In principle, the maximum time units required for multi-stage RSA to finish are equal to the number of stages in the algorithm. However, in practice, some members will be labeled negative at earlier stages and can therefore exit the process sooner. Let $w_l$ denote the number of time units that members spend in stage $l$ in a multi-stage RSA. In the first stage, the entire population of $n$ members is subject to joint tests, meaning that all members must remain in the algorithm for $w_1$ time units as all states are ambiguous at this point. In subsequent stages, only the suspected members proceed to the next stage where they must wait for an additional $w_{l+1}$ time units. This process continues recursively until the final stage, by which point only a small fraction of the original population remains. Let $W$ be a random variable representing the duration of any multi-stage RSA, with an unknown distribution $F_W$. Similar to the ENT, the goal is to quantify the expected duration $E(W)$. The resulting expression for $E(W)$ is formalized in Theorem \ref{theorem:dur}. A proof is given in Appendix \ref{app:theorem:dur}.

\begin{theorem} \label{theorem:dur}
If the population size is sufficiently large $n \to \infty$, GT with adaptive, probabilistic and conservative multi-stage RSA can be completed in
    \begin{equation*}
        E(W) \stackrel{a}{\sim} n \left( w_1 + \sum_{l=2}^{k+1} w_l \left(p + q(1-q^{s_{l-1}-1})^{r_{l-1}}\right) \right)
    \end{equation*}  
time on average.
\end{theorem}

Theorem \ref{theorem:dur} implies that the expected duration depends on the size of the population $n$, the time units each member spends in the stages $w_l$, the constant members per test $s_l$, and the constant number of joint tests per member $r_l$. Furthermore, note that I explicitly assume that the joint tests $r_l$ can be conducted in parallel. However, premultiplying the expression by an additional $r_l$ in each stage yields a similar result which quantifies $E(W)$ when the joint tests can only be conducted sequentially. In principle, the expected duration can be utilized as an additional performance criterion in the evaluation of a multi-stage RSA. The general approach is to optimize the ENT, quantify the $s_l$ and $r_l$ in the minimum and then compute the $E(W)$ based on Theorem \ref{theorem:dur}. The ENT can always be compared together with the expected duration which allows for a quantification of the trade-off between the duration and the tests.

\section{Simulation and Performance Evaluation for the Expected Duration} \label{sec:app2}
\begin{figure}[!htpb]
    \centering
    \includegraphics[width = \textwidth]{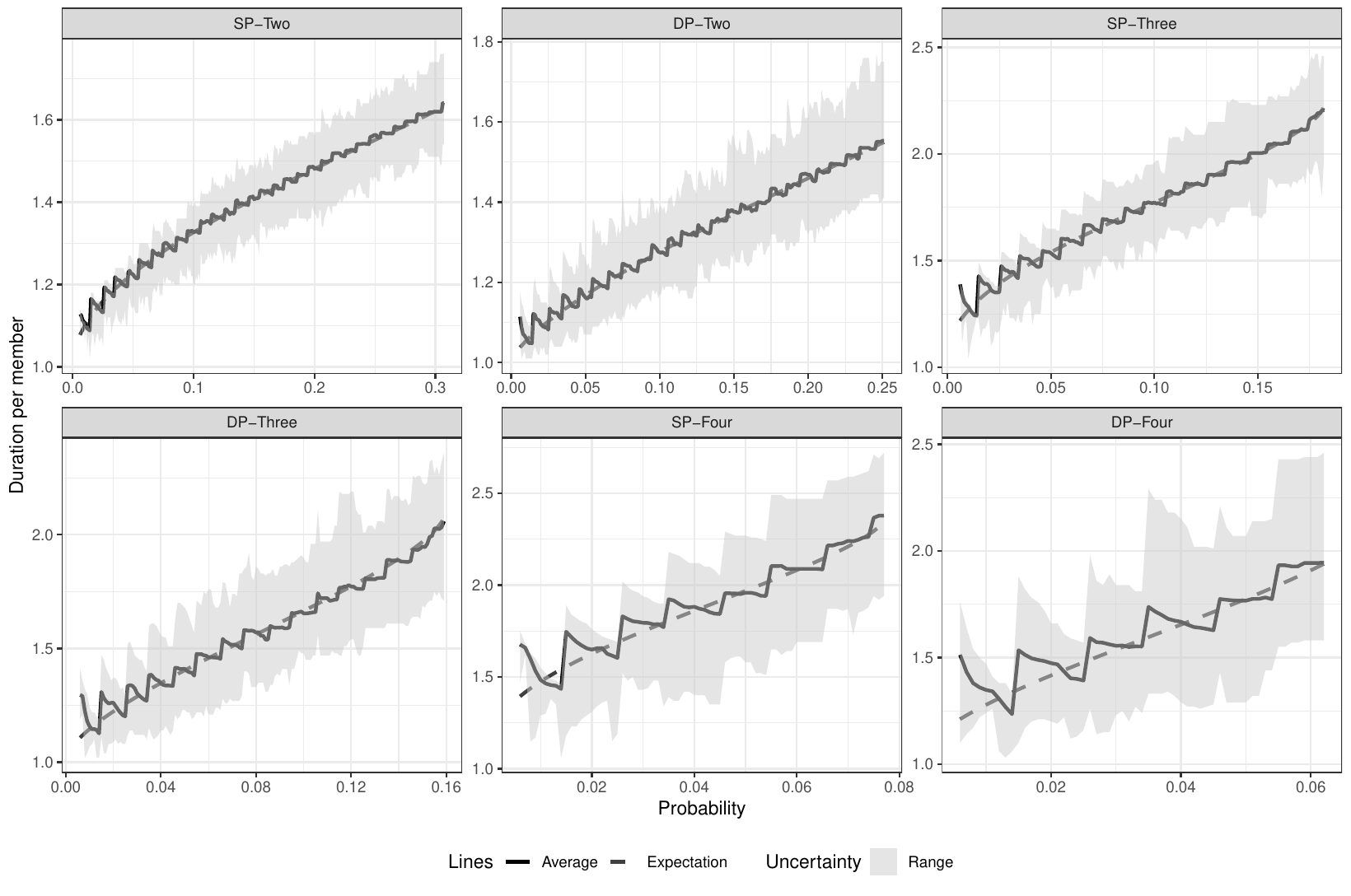} \hfill
    \caption{Comparison of expected duration per member pictured as black dashed lines and average duration per member picture as gray straight lines for admissible probabilities $p$ and a small population size $n = 100$.}
    \label{fig:sim3}
\end{figure}

\begin{figure}[!htpb]
    \centering
    \includegraphics[width = \textwidth]{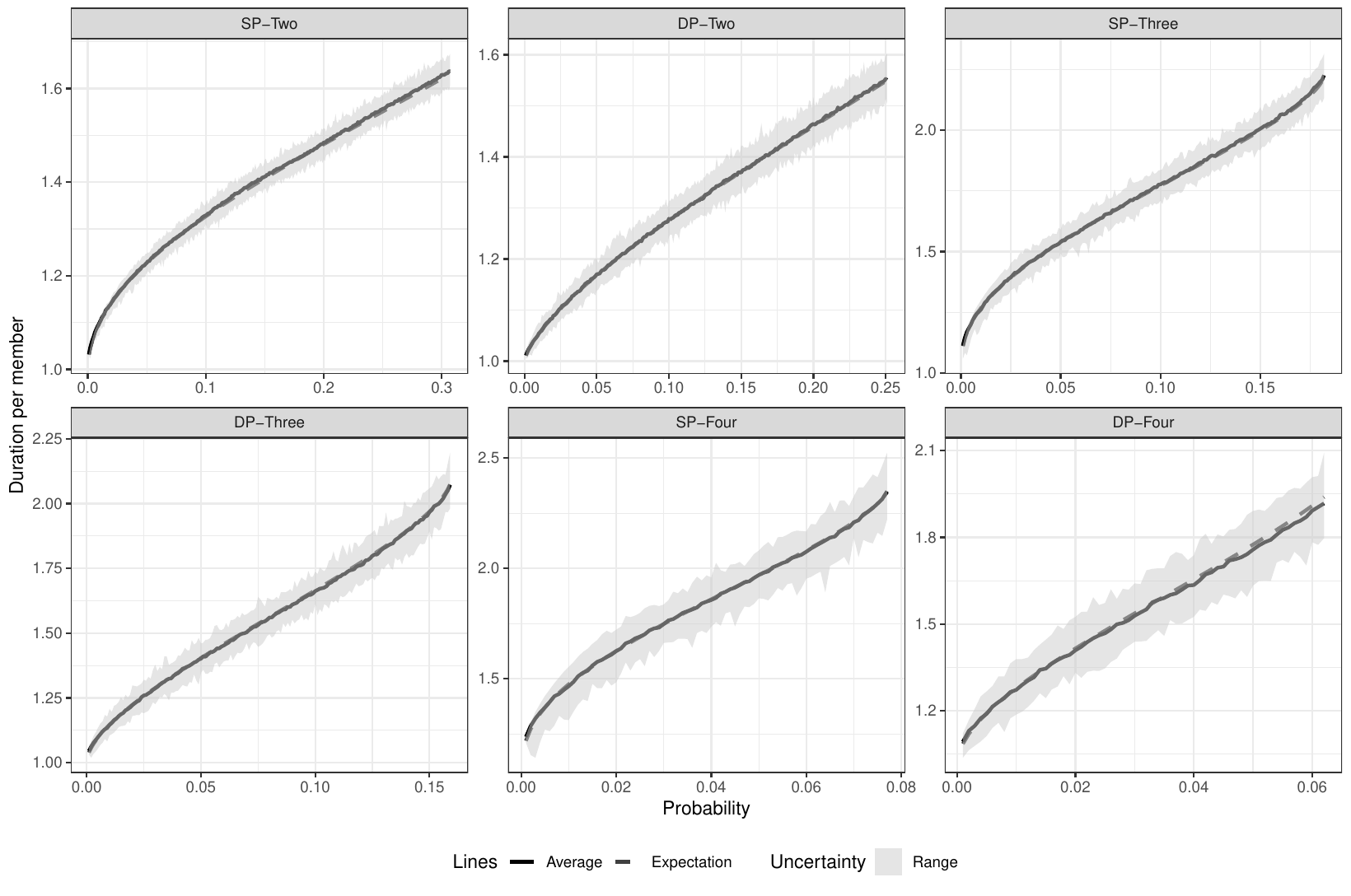} \hfill
    \caption{Comparison of expected duration per member pictured as black dashed lines and average duration per member picture as gray straight lines for admissible probabilities $p$ and a finite population size $n = 1000$.}
    \label{fig:sim4}
\end{figure}

The results for the expected duration are presented in the following section. I begin by providing simulation studies for the expected duration with a small population size of $n = 100$. The results are shown in Figure \ref{fig:sim3}.
Overall, the behavior of the average duration per member closely mirrors that of the ATM. The interpretation is also analogous. For small population sizes, the number of possible group configurations with defective members is limited. Consequently, in simulations, this small population size combined with low probabilities can lead to significant underestimation or overestimation of the expected duration per member. Moreover, the uncertainty bands are quite large for all GT algorithms, indicating a broader spread of duration values in the simulations. In Figure \ref{fig:sim4}, the results for a larger population size of $n = 1000$ are presented.
As with the ETM, the results show that the average and expected duration per member converge and are almost identical for the larger population size. A small underestimation is only observed for DP-Four in combination with high probability values. The abrupt fluctuations seen with $n = 100$ have also disappeared supporting the asymptotic equivalence in Theorem \ref{theorem:dur}. Uncertainty is now much lower across all GT algorithms, indicating the proper functionality of the simulated algorithms.

Finally, the MAPE for both settings across different probability intervals is evaluated. The results are shown in Table \ref{tab:mape2}.
\begin{table}[!htpb]
\centering
\caption{\label{tab:mape2}  Mean absolute percentage error (MAPE) results for the expected duration with per member very small $n = 100$ and finite $n = 1000$ population size for admissible probability intervals.}
\scriptsize
\setlength{\tabcolsep}{2pt} 
\begin{tabular}{l*{6}{c}}
\toprule
\multirow{2}{*}{Algorithm} & \multicolumn{3}{c}{$n = 100$} & \multicolumn{3}{c}{$n = 1000$} \\
\cmidrule(lr){2-4} \cmidrule(lr){5-7}
& $p < 0.077$ & $p \in (0.077, 0.182)$ & $p > 0.182$ & $p < 0.077$ & $p \in (0.077, 0.182)$ & $p > 0.182$ \\
\midrule
SP-Two & 1.23\% & 0.524\% & 0.371\% & 0.0865\% & 0.240\% & 0.300\% \\
DP-Two & 1.42\% & 0.649\% & 0.501\% & 0.0842\% & 0.106\% & 0.192\% \\
SP-Three & 2.45\% & 0.679\% & - & 0.145\% & 0.287\% & - \\
DP-Three & 2.86\% & 1.12\% & - & 0.194\% & 0.277\% & -\\
SP-Four & 3.26\% & - & - & 0.220\% & - & -\\
DP-Four & 4.33\% & - & - & 0.582\% & - & - \\
\bottomrule
\end{tabular}
\end{table}
The results are again similar to those for the ETM. For small $n = 100$ and low $p < 0.077$ , the MAPE exhibits the largest values. It decreases rapidly as $p$ increases, eventually reaching values around 1\%. For $n = 1000$, the MAPE is consistently around 0.75\%, indicating a low deviation between the expected and average duration per member. Across all $p$ values, the two-stage SP has the closest values of the average to the expected duration per member. Increasing the number of stages or the number of joint tests leads to a larger MAPE, simply because the GT algorithms become more complex. As before, GT algorithms are only applicable within certain probability intervals. Four-stage algorithms are only applicable for $p < 0.077$, three-stage algorithms up to $p = 0.182$, and two-stage algorithms up to $p = 0.35$. Beyond this, individual testing becomes the only viable option, as the probability of a member being defective is so high that it is no longer possible to partition the population into negative groups. 

\begin{figure}[!htpb] 
    \centering
    \includegraphics[width = \textwidth]{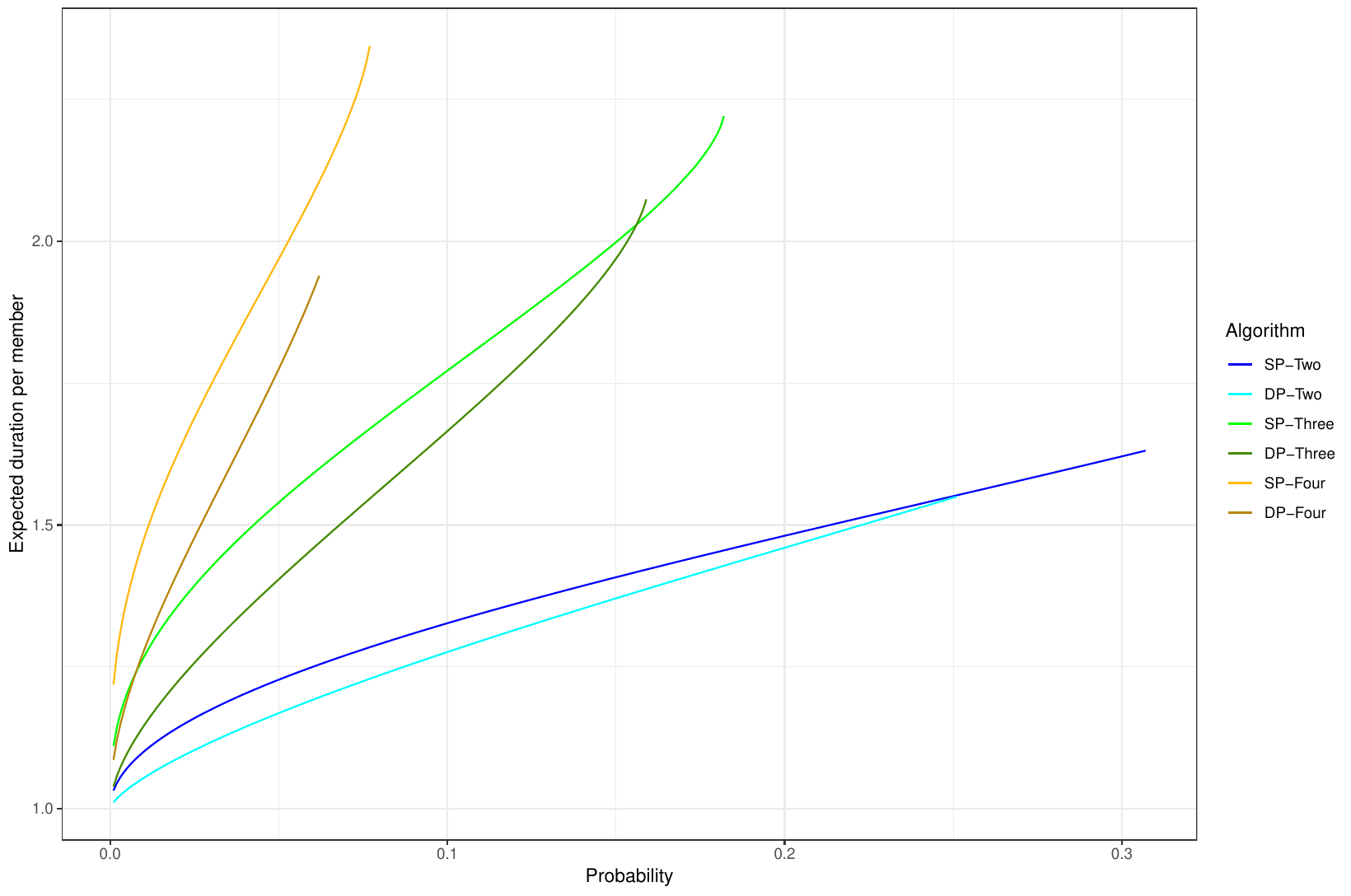} \hfill
    \caption{Performance evaluation of expected duration per member for eligible probabilities $p$. Each line represents the progress of the expected duration per member for the complete probability range.}
    \label{fig:per3}
\end{figure}

In the final step, the performance of different GT algorithms is compared by evaluating the expected duration per member across the full range of admissible probabilities $p$. The results are shown in Figure \ref{fig:per3}.
It can be concluded that two-stage SP and DP have the lowest expected duration per member across the entire probability range. This is not surprising, as both are two-stage GT algorithms. They are followed by the three-stage SP and DP. Interestingly, the results suggest that the three-stage DP combines a low ETM with a shorter duration compared to the three-stage SP. The highest expected duration per member is observed for the four-stage GT algorithms. Similar to the three-stage DP, four-stage DP algorithm exhibits a shorter overall duration compared to the four-stage SP counterpart. In summary, the results indicate that multi-stage DP consistently outperform their multi-stage SP counterparts, while also providing a lower ETM over certain probability intervals. However, Theorem \ref{theorem:dur} implicitly assumes that the joint tests in the first stage can all be conducted in parallel. If a parallel testing procedure is not feasible, the results for the expected duration changes towards classical multi-stage 1SA.

\section{Collection of Proofs} \label{app:proofs}
\subsection{Proof of Lemma \ref{lm:bin}} \label{app:lm_bin}
Without loss of generality, consider a population of size $n \to \infty$ and a group size $s$ which is fixed and $s = o(\sqrt{n})$. Then, from Stirling's approximation $n! \stackrel{a}{\sim} \sqrt{2 \pi n} \left(\frac{n}{e}\right)^n$ and first order Taylor approximation of the logarithm $\log\left(1-\frac{s}{n}\right) \approx -\frac{s}n{}$, it follows
\begin{align*}
    \binom{n}{s} &= \frac{n!}{s!(n-s)!} \\
                 & \stackrel{a}{\sim}  \frac{\sqrt{2 \pi n} \left(\frac{n}{e}\right)^n}{s! \sqrt{2 \pi (n-s)} \left(\frac{(n-s)}{e}\right)^{(n-s)}} \\
                 &= \frac{\sqrt{ n} \left(\frac{n}{e}\right)^n}{s! \sqrt{(n-s)} \left(\frac{(n-s)}{e}\right)^{(n-s)}} \\
                 &= \frac{\sqrt{n}n^ne^{-n}}{s!\sqrt{(n-s)} (n-s)^{(n-s)}e^{-(n-s)}} \\
                 &= \frac{\sqrt{n}}{\sqrt{n-s}} \frac{e^{-s} n^n}{s!(n-s)^{(n-s)}} \\
                 &= \frac{n^s}{s!} e^{-s} \left(1-\frac{s}{n}\right)^{-n} \left(1-\frac{s}{n}\right)^{s-\frac{1}{2}} \\
                 &\approx \frac{n^s}{s!} \left(1-\frac{s}{n}\right)^{s-\frac{1}{2}} \\
                 &\stackrel{a}{\sim} \frac{n^s}{s!} 
\end{align*}
\qed

\subsection{Proof of Lemma \ref{lm:groups}} \label{app:lm_groups}
Without loss of generality, consider a population of size $n$ where each member $i$ is i.i.d. defective with probability $P(X_i = 1) = p$. Then $P(X_i = 0) = q = 1 - p$ is the probability of a member being non-defective. Divide this population into disjoint groups $G$ of size $s$. Pick one group $g \in G$ out of the set of all groups. The probability that any single member $i$ of a group $g$ is non-defective is again $P(X_i = 0) = q$ because the Bernoulli trials are assumed to be identical and independent and $g \subset n$. Since the state of the members of the group are assumed to be independent of each other, the probability that all $s$ members are non-defective is the product of the individual probabilities of their states, given by $P\left(\sum_{i = 1}^s X_i = 0\right) = P(\text{g is negative}) = q^s$. Conversely, the probability that a group $g$ is positive, that is, it contains at least one defective member, is  $P\left(\sum_{i = 1}^s X_i  \geq 1\right) =  P(\text{g is positive}) = (1 - q^s)$, which is the complement of the event that all members are non-defective.\qed

\subsection{Proof of Lemma \ref{lm:sus}} \label{app:lm_sus}
Without loss of generality, consider a population of size $n$ where each member $i$ is defective with probability $P(X_i = 1) = p$ and non-defective with probability $P(X_i = 0) = q = 1 - p$ with a total of fixed $d$ defective members. Arrange the members in the population randomly $r$ times. In each random permutation, divide the population into disjoint groups $G$ of size $s$. Pick one group $g \in G$ and let one member $i^*$ in this group be non-defective. Then there are $\binom{n-1}{s-1}$ possible ways to fill the remaining $s-1$ spots in the group with other members from the population. If the group is negative, then, from Lemma \ref{lm:groups}, the remaining $s-1$ members must all be non-defective. The number of ways to fill these $s-1$ with non-defective members from the $n-d-1$ non-defective members is $\binom{n-d-1}{s-1}$. The probability that the group without one non-defective member is negative is then asymptotically approximated by Lemma \ref{lm:bin} if $n$ (and thus $n-1 $ and $n-d-1$) is much larger than $s$ (and thus $s-1$) as
\begin{align*}
P(g \setminus \{i^*\} \ \text{is negative}) &= \frac{\binom{n-d-1}{s-1}}{\binom{n-1}{s-1}} \\ 
&\stackrel{a}{\sim} \left(\frac{\frac{(n-d-1)^{s-1}}{(s-1)!}}{\frac{(n-1)^{s-1}}{(s-1)!}}\right)\\
&= \left(\frac{n-d-1}{n-1}\right)^{s-1} \\
&= \left(1 - \frac{d}{n-1}\right)^{s-1} \\
&\approx (1-p)^{s-1} \\
&= q^{s-1}.
\end{align*}
The complement which is the probability that the group without one defective member is positive is then $P(g \setminus \{i^*\} \ \text{is positive}) = 1 - q^{s-1}$. Since each joint test $r$ is conducted independently, the probability that a member is suspected, meaning the state is currently ambiguous, is the probability that the member is either defective or non-defective but part of a positive group in $r$ joint tests. This can be expressed as $P(Y_i = 1) = p + q(1 - q^{s-1})^r$. Consequently, the probability that the state is not ambiguous is given by $P(Y_i = 0) = 1 - (p + q(1 - q^{s-1})^r)$.
\qed

\subsection{Proof of Theorem \ref{theorem:cent}} \label{app:theorem_cent}
Consider a population of size $n$ where each member $i$ is defective with probability $P(X_i = 1) = p$ and non-defective with probability $P(X_i = 0) = q = 1 - p$. In total, there are $d$ defective members in the population. Set $l = 1$ and arrange the members in the population randomly $r_1$ times, then the population $n$ is grouped into $\frac{n}{s_1}$ for a total of $r_1$. Thus, the total number of tests in the first stage is equal to the number of groups multiplied by the number of joint tests. Hence, $E(T_1)$ is given by
\begin{equation*}
    E(T_1) = T_1 = \frac{nr_1}{s_1}.
\end{equation*}
Let $n_1$ denote the number of suspected members after the first stage. In principle, $n_1$ is a random variable, since it can be interpreted as the sum of ambiguous states of the members $n_1 = \sum_{i = 1}^n Y_i^{(1)}$. The probability distribution is then given by 
\begin{equation*}
    f_1(n_1 \mid n, p_1) = \binom{n}{n_1} p_1^{n_1} (1-p_1)^{n-n_1}.
\end{equation*}
From Lemma \ref{lm:sus}, the probability that a member is suspected is quantified via $P(Y_{i}^{(1)} = 1) = p_1 = p + q(1 - q^{s_1-1})^{r_1}$. Consequently, the expected number of suspected members in stage one is
\begin{equation*}
    E(n_1) = np_1 \stackrel{a}{\sim} n \left(p + q(1-q^{s_1-1})^{r_1}\right).
\end{equation*}
Set $l = 2$ and arrange the suspected members in the population $n_1$ randomly $r_2$ times. Then regroup these suspected members into groups of size $s_2$ for a total of $r_2$, thereby resulting in a number of $T_2 = \frac{r_2}{s_2} n_1$ tests in the second stage. Hence, $E(T_2)$ is given by 
\begin{equation*}
    E(T_2) = E\left(\frac{r_2}{s_2} n_1\right) \stackrel{a}{\sim} \frac{r_2}{s_2} n \left(p + q(1-q^{s_1-1})^{r_1}\right).
\end{equation*}
For any stage $l \in \{3, \dots, k\}$, denote $n_{l-1}$ the number of suspected members after stage $l - 1$. This implies again that $n_{l-1} = \sum_{i = 1}^{n_{l-2}} Y_i^{(l-1)}$ is a random variable which probability distribution is given by 
\begin{equation*}
    f_{l-1}(n_{l - 1} \mid g_{l-2}, p_{l - 1 \mid l-2}) = \binom{n_{l-2}}{n_{l - 1}} p_{l - 1 \mid l-2}^{n_{l - 1}} (1-p_{l - 1 \mid l-2})^{n_{l-2}-n_{l -1}},
\end{equation*}
where $p_{l - 1 \mid l-2}$ is the probability that a member is suspected in stage $l - 1$ given that the member was also suspected in the previous stage. This probability can be derived using conditional probabilities. Without loss of generality, let $A$ be the event that a member is suspected in stage $l-2$, and let $B$ be the event that the member is suspected in stage $l-1$, given that the member was suspected in stage $l-2$. Then following relationship holds
\begin{equation*}
    P(A \cap B) = P(B) = p_{l-1},
\end{equation*} 
where $p_{l-1} = p + q(1-q^{s_{l-1}-1})^{r_{l-1}}$ obtained in the same manner as $p_1$ due to Lemma \ref{lm:sus}. Since $n_{l-1} < n_{l-2}$ always holds, a member can only be suspected in stage $l-1$, if it was also already suspected in stage $l-2$. Thus, $P(A \cap B)$ is simply the probability of being a suspected member nested under previously suspected members $P(B)$. The conditional probability $p_{l -1 \mid l-2}$  is then given by
 \begin{equation*}
   p_{l -1 \mid l-2} = \frac{P(A \cap B)}{P(A)} = \frac{p_{l-1}}{p_{l - 2}}.
\end{equation*}
Therefore, the expected number of suspected members in stage $l$ is
\begin{align*}
    E(n_{l-1}) &= E\left(\prod_{j=1}^{l - 1} n_j\right) \\
            &= n \left(p_1 \prod_{j=2}^{l - 1} p_{j \mid j -1}\right) \\
            &= n \left(p_1 \prod_{j=2}^{l - 1} \frac{p_l}{p_{l - 1}}\right) \\
            &\stackrel{a}{\sim} n \left(p + q(1-q^{s_{l-1}-1})^{r_{l - 1}}\right).
\end{align*}
Arrange these suspected members in the population $n_{l-1}$ randomly $r_l$ times and regroup them into groups of size $s_l$ for a total of $r_l$. The total number of test $T_l$ in stage $l$ is then given by $T_l = \frac{r_l}{s_l} n_{l-1}$. Thus, the expected number of tests $E(T_l)$ is
\begin{equation*}
    E(T_l) = E\left(\frac{r_l}{s_l} n_{l-1}\right) \stackrel{a}{\sim} \frac{r_l}{s_l} n \left(p + q(1-q^{s_{l - 1}-1})^{r_{l - 1}}\right).
\end{equation*}
In the final stage $k + 1$, individual retesting of suspected members is necessary. Let $n_k$ denote the number of members whose state is still ambiguous after $k$ previous stages of joint group tests. The probability distribution for the number of remaining suspected members is given by
\begin{equation*}
    f_{k}(n_k \mid n_{k - 1}, p_{k \mid k - 1}) = \binom{n_{k - 1}}{n_{k }} p_{k \mid k - 1}^{n_{k}} (1-p_{k \mid k-1})^{n_{k -1} - n_{k}}.
\end{equation*}
In this final stage, the expected number of suspected members is equal to the ENT $E(T_{k+1})$, since joint group tests are no longer applied. This gives
\begin{equation*}
    E(T_{k+1}) = E(n_k) \stackrel{a}{\sim} n \left(p + q(1-q^{s_{k}-1})^{r_{k}}\right).
\end{equation*}
Combining the results from all stages, summing up the ENT obtained in each stage and utilizing the linearity of expectation:
\begin{align*}
    E(T) &= E(T_1) + E(T_2) + \sum_{l=3}^{k} E(T_l) + E(T_{k + 1}) \\
          &\stackrel{a}{\sim} \frac{nr_1}{s_1} + \frac{r_2}{s_2} n \left(p + q(1-q^{s_1-1})^{r_1}\right) + \sum_{l=3}^{k} \frac{r_l}{s_l} n \left(p + q(1-q^{s_{l - 1}-1})^{r_{l - 1}}\right) \\
          &\quad + n \left(p + q(1-q^{s_{k}-1})^{r_{k}}\right) \\
          &= \frac{nr_1}{s_1} + \sum_{l=2}^{k} \frac{r_l}{s_l} n \left(p + q(1-q^{s_{l - 1}-1})^{r_{l - 1}}\right) + n \left(p + q(1-q^{s_{k}-1})^{r_{k}}\right) \\
          &= n \left( \frac{r_1}{s_1} + \sum_{l=2}^{k} \frac{r_l}{s_l} \left(p + q(1-q^{s_{l-1}-1})^{r_{l-1}}\right) + p + q(1-q^{s_k-1})^{r_k} \right).
\end{align*}
\qed

\subsection{Proof of Theorem \ref{theorem:dur}} \label{app:theorem:dur}
Consider a population of size $n$ where each member $i$ is defective with probability $P(X_i = 1) = p$ and non-defective with probability $P(X_i = 0) = q = 1 - p$. In total, there are $d$ defective members in the population. In each stage $l$, the population is arranged randomly for $r_l$ times and then grouped into groups of size $s_l$. Set $l = 1$ and assume that the duration of this stage is $w_1$ time units per member. By construction, the state of all members is ambiguous, which in turn indicates that all members remain for the entire duration $W_1$ in the GT procedure. Thus, the expected duration $E(W_1)$ in the first stage is given by
\begin{equation*}
     E(W_1) = W_1 = w_1n.
\end{equation*}
Let $n_1$ denote the number of suspected members after the first stage. As in Theorem \ref{theorem:cent}, $n_1$ is a random variable, since it can be interpreted as the sum of ambiguous states of the members $n_1 = \sum_{i = 1}^n Y_i^{(1)}$. The expected number of suspected members is $E(n_1) \stackrel{a}{\sim} n \left(p + q(1-q^{s_1-1})^{r_1}\right)$ . Set $l = 2$. Each suspected member in the population remains in the GT procedure for an additional $w_2$ time units. Therefore, the expected duration $E(W_2)$ in stage two is
 \begin{equation*}
     E(W_2) = E(w_2n_1) \stackrel{a}{\sim} w_2 n \left(p + q(1-q^{s_1-1})^{r_1}\right).
\end{equation*}
For any stage $l \in \{3, \dots, k\}$, denote $n_{l-1}$ the number of suspected members after stage $l - 1$. The expected number of suspected members in stage $l$ is given by $E(n_{l-1}) \stackrel{a}{\sim} n \left(p + q(1-q^{s_{l-1}-1})^{r_{l-1}}\right)$. Each suspected member in the population remains in the GT procedure for an additional $w_l$  time units. Thus, the expected duration $E(W_l)$ in stage $l$ is
\begin{equation*}
     E(W_l) = E(w_ln_{l-1}) \stackrel{a}{\sim} w_l n \left(p + q(1-q^{s_{l-1}-1})^{r_{l-1}}\right).
\end{equation*}
In the final stage $k + 1$, individual retesting of suspected members is necessary. Let $n_k$ denote the number of members whose state is still ambiguous after $k$ previous stages of joint group tests. The expected number of suspected members is $E(n_k) \stackrel{a}{\sim} n \left(p + q(1-q^{s_k-1})^{r_k}\right)$ . The duration per member is $w_{k+1}$ time units, and the expected duration $E(W_{k+1})$ is given by
\begin{equation*}
     E(W_{k+1}) = E(w_{k+1}n_k) \stackrel{a}{\sim} w_{k+1} n \left(p + q(1-q^{s_k-1})^{r_k}\right).
\end{equation*}
Finally, combining the expected duration from all stages and utilizing the linearity of expectation yields
\begin{align*}
      E(W) &=E(W_1) + E(W_2) + \sum_{l=3}^{k} E(W_l) + E(W_{k+1}) \\
      &\stackrel{a}{\sim} w_1n + w_2 n \left(p + q(1-q^{s_1-1})^{r_1}\right) + \sum_{l=3}^{k} w_l n \left(p + q(1-q^{s_{l-1}-1})^{r_{l-1}}\right) \\
      &\quad + w_{k+1} n \left(p + q(1-q^{s_k-1})^{r_k}\right) \\
       &= n \left( w_1 + \sum_{l=2}^{k+1} w_l \left(p + q(1-q^{s_{l-1}-1})^{r_{l-1}}\right) \right).
\end{align*}
\qed

\end{appendices}

\bibliography{sn-bibliography}

\end{document}